\documentclass[12pt, aps, pra,nofootinbib]{revtex4-1}
\usepackage{epsfig}

\begin{document}

\begin{center} 
{\Large {\bf Photino and Gluino Production in SQED and SQCD}}
\end{center}

\vspace{5mm}

\begin{center}
D.B. Espindola$^a$, M. C. Rodriguez$^b$ and C. Brenner Mariotto$^b$  \\
$^a${\it Universidade Federal do Rio Grande do Sul \\
Instituto de F\'\i sica - IF-UFRGS \\
Av. Bento Gon\c calves, 9500 \\
Caixa Postal 15051, Cep:91501-970, \\
Porto Alegre, RS \\
Brazil} \\
$^b${\it Universidade Federal do Rio Grande - FURG \\
Instituto de Matem\'atica, Estat\'\i stica e F\'\i sica - IMEF \\
Av. It\'alia, km 8, Campus Carreiros \\
Caixa Postal 474, Cep:96200-970, \\
Rio Grande, RS \\
Brazil} 
\vspace{5mm}
\\
\end{center}

\begin{abstract}
  \vspace{3mm} This review starts out with a brief history of the
  photino-gluino phenomenology. Next, it describes difficulties
  encountered in the construction of Feynman amplitudes when dealing
  with Majorana fermions and outlines a procedure that circumvents this
  difficulty by means of a prescription based on a well-defined fermion
  flow without explicit charge-conjugation matrices at the vertex
  equations. Finally, as an illustration, it calculates the cross
  section for the production of photinos and gluinos.
\end{abstract}

PACS   numbers: 12.60.-i %Models beyond the standard model
12.60.Jv %Supersymmetric Model
13.85.Lg %Total cross section
 
\maketitle
%\vspace{5mm}

\section{Introduction}

Supersymmetry (SUSY) was introduced more than 30 years ago, in
independent theoretical papers by Golfand and Likhtman~\cite{gl},
Volkov and Akulov~\cite{va}, and Wess and Zumino~\cite{wz}. In the
first of these articles \cite{gl}, the authors found the superextension of the
Poincar\'e algebra and constructed the first four-dimensional field
theory with supersymmetry, the (massive) supersymmetric quantum
eletrodynamics (SQED).

The particle content of SQED is given by the electron (fermion),
positron (antifermion), which are Dirac fermions \cite{Dirac}, the
photon, and the spin-$0$ partners of the electron and the positron,
the so-called selectron and spositron (sfermions). Also included in
the model is the photino, the spin-$(1/2)$ superpartner of the
photon. The photon and the photino are introduced in the same vector
superfield, therefore the photino must be a Majorana fermion
\cite{Majorana}. In this theory both the $R$-Parity and the chirality
are conserved and sparticles appear in pairs at any vertex, see
\cite{wb,dress,MullerKirsten:1986cw,bailin,tata,ait1,Srednicki:2004hg}.

The history of the Minimal Supersymmetric Standard Model (MSSM), which
was constructed in 1975 \cite{R,ssm,grav}, can be found in
Ref.~\onlinecite{Fayet:2001xk,Rodriguez:2009cd}.  In the early days of
the model, the photino, which is denoted by the symbol $\tilde \gamma$,
was called the``photon-neutrino" and envisioned as a fundamental
particle, expected to be stable \cite{Goldberg:1983nd} and, at least
at the classical level, massless \cite{Fayet:1979yb}.

Within the context of supersymmetric field theories, two-component
Weyl-van der Waerden fermions~\cite{vdWaerden1} enter naturally, due
to the spinorial nature of the symmetry generators themselves and the
holomorphic structure of the superpotential. Using the two-component
spinors we can introduce the helicity formalism, in which the
individual amplitudes are computed analytically in terms of Lorentz
scalar invariants, i.~e., of complex numbers that can readily be
computed. It is then a simple numerical task to sum all the
contributing amplitudes and compute the square of the complex
magnitude of the resulting sum
\cite{Haber:1994pe,Martin:tasi,dreiner1}.\footnote{Ref.~\cite{dreiner1}
  discusses the history of the spinors and the techniques
  in the two-component spinor formalism and provides a complete set of
  Feynman rules for fermions in two-component spinor notation.}

As a simple example, consider Bhabha scattering $e^{-}e^{+}\rightarrow
e^{-}e^{+}$ \cite{bhabha} in QED. We denote the initial-state momenta
and helicities of the electron and positron ($p_{1},\lambda_{1}$) and
($p_{2},\lambda_{2}$) , respectively, and the final-state momenta and
helicities of the electron and positron ($p_{3},\lambda_{3}$) and
($p_{4},\lambda_{4}$), respectively.  In the formalism
of four-component Dirac spinors, the amplitude for the $s$ channel is given by the equality
\begin{equation}
{\cal M}= \frac{e^{2}}{s}\bar{u}(p_{3},\lambda_{3})\gamma^{m}v(p_{4},\lambda_{4}) 
\bar{v}(p_{2},\lambda_{2})\gamma_{m}v(p_{1},\lambda_{1}).
\end{equation}

For the same process, in the two-spinor formalism the amplitude is given
by the expression \cite{Haber:1994pe,Martin:tasi,dreiner1}:
\begin{eqnarray}
\imath {\cal M}&=& \left( \frac{\imath g^{mn}}{s} \right) \left[
( \imath e x_{1}\sigma_{m}y^{\dagger}_{2})
(- \imath e y_{3}\sigma_{n}x^{\dagger}_{4})+ 
( \imath e y^{\dagger}_{1}\bar{\sigma}_{n}x_{2})
(- \imath e y_{3}\sigma_{n}x^{\dagger}_{4}) \right. \nonumber \\
&+& \left. ( \imath e x_{1}\sigma_{m}y^{\dagger}_{2})
(- \imath e x^{\dagger}_{3}\bar{\sigma}_{n}y_{4})+ 
( \imath e y^{\dagger}_{1}\bar{\sigma}_{n}x_{2})
(- \imath e x^{\dagger}_{3}\bar{\sigma}_{n}y_{4})
\right] \,.
\end{eqnarray} 
For a more extensive discussion of this formalism and other useful
examples,  see Ref.~\cite{dreiner1}.

Parity-conserving theories such as QED and QCD are well-suited to the
four-component fermion methods. The latter being more broadly known
than the two-component methods, our discussion will be based on
Majorana-(Dirac) four-component spinors.

Early in its history, the gluino, which is denoted $\tilde g$, was
called the ``photonic neutrino" and viewed as a massless particle,
since it was difficult to generate a sizeable mass for it. The role
and interactions of this fermion partner of the gluon, which like the
photino is a Majorana fermion, are directly related to the
properties of the supersymmetric QCD (SQCD) \cite{wb,dress}.  In the
early days, the existence of relatively light
``$R$-hadrons"\footnote{Particle made of quarks, antiquarks and
  gluinos.}~\cite{ff,ff2} was therefore expected. Today we know that a
direct gaugino mass, symbolized $m_{1/2}$, comes from supergravity
\cite{Cremmer:1982vy}, or from radiative corrections using messenger
quarks. Both mechanisms yield sufficiently high masses for the gluino.

The phenomenological studies of the photinos started in 1979, when
Fayet \cite{Fayet:1979yb} studied the interaction between
photinos\footnote{In the MSSM context the photino interacts more
  weakly than the neutrino.} and matter for massless photinos and
obtained the following cross section:
\begin{equation}
\sigma ( \tilde{\gamma}+e^{-} \to \tilde{\gamma}+e^{-})= \frac{4G^{2}_{F}m_{e}E}{6 \pi} 
\left( \frac{4M^{2}_{W} \sin^{2} \theta_{W}}{m^{2}_{s_{e}}} \right)^{2}\,\,,
\end{equation}
where $m_{s_{e}}$ is the mass of the slepton. See
Ref.~\cite{Rodriguez:2009cd} for the early notation.

Later, in 1982,  Fayet \cite{Fayet:1982ky} studied the photino
production from $e^{-}e^{+}$, still in the case of massless photinos,
with the following result:
\begin{equation}
\frac{d \sigma}{d \Omega}(e^{-}e^{+} \to \tilde{\gamma}\tilde{\gamma})= \frac{\alpha^{2}s}{16} \left[
\frac{(1- \cos^{2} \theta)^{2}}{\left( m^{2}_{s_{e}}+ \frac{s}{2}(1- \cos^{2} \theta) \right)^{2}}+
\frac{(1+ \cos^{2} \theta)^{2}}{\left( m^{2}_{t_{e}}+ \frac{s}{2}(1+ \cos^{2} \theta) \right)^{2}}
\right],
\label{fayetresult}
\end{equation}
where $\theta$ is the angle between the photino and the incoming
electron, and $s$ is the usual Mandelstam variable. 

A later computation considered a massive photino \cite{kane,Dawson}.
The mass $m_{\tilde{\gamma}}$ acquired phenomenological importance
because it was related to the scale of supersymmetry breaking
\cite{Goldberg:1983nd}, to which, more recently, the mass of the
gluino has also been related. With a massive photino, the total cross
section is \cite{Fayet:1982ky}
\begin{eqnarray}
  \sigma(e^{-}e^{+} \to \tilde{\gamma}\tilde{\gamma})= 
  \frac{2  \pi \alpha^{2}s}{3m^{4}_{s_{e}}}.
\end{eqnarray}

Fayet also analyzed the processes $e^{-}e^{+} \to \gamma \nu \bar{\nu}$ and 
$e^{-}e^{+} \to \gamma \tilde{\gamma}\tilde{\gamma}$ to obtain the
following total cross sections (in pb) \cite{Fayet:1982ky}:
\begin{eqnarray}
\sigma(e^{-}e^{+} \to \gamma \nu \bar{\nu})&\approx&2.6 \cdot 10^{-2} \frac{s}{(40 \mbox{GeV})^{2}}, \nonumber \\
\sigma(e^{-}e^{+} \to \gamma \tilde{\gamma}\tilde{\gamma})&=&18 \left( \frac{m_{s_{e}}}{40 \frac{\mbox{GeV}}{c^{2}}} \right)^{-4} \frac{s}{(40 \mbox{GeV})^{2}}.
\end{eqnarray}

Only in 1984 were the photino and selectron masses included in these
processes \cite{Kobayashi:1984wu,Grassie:1983kq,Ware:1984kq}. At the
time, it was thought that these reactions might define a useful
signature of SUSY and the experiments, important limits on the
photino and selectron masses. Two years later the reactions
$e^{-}e^{+} \to \gamma \nu \bar{\nu}$ and $e^{-}e^{+} \to \gamma
\tilde{\gamma}\tilde{\gamma}$ were accurately analyzed, and the
results showed that the latter process has larger cross section than
the former (see Fig.~2 in Ref.~\onlinecite{Ware:1984kq}), a finding
applicable only to the ``lower'' selectron masses.  It was also found
that the processes with polarized and unpolarized beams place strong
limits on the $m_{\tilde{\gamma}} \times m_{\tilde{e}}$ plane (see
Fig.~4 in Ref.~\onlinecite{Ware:1984kq}) \cite{Bento:1985in}.  A
photino production in this channel was analyzed in detail in
Ref.~\onlinecite{pandita}.

Low-mass weakly interacting particles (photinos, neutrinos, axions,
etc.) are produced in hot astrophysical plasmas and can therefore
transport energy out of stars. The possible astrophysical consequences
of ``light" photinos and gluinos, for which the main photino production
channel is the subprocess $gg \to \tilde{g}\tilde{g}$ followed by a
gluino decay $\tilde{g} \to \tilde{\gamma}\bar{q}q$, were discussed in
Refs.~\onlinecite{Goldberg:1983nd,Berezinsky:1986ty}.

It has been generally assumed that for small gaugino masses the
photino is an approximate eigenstate, an assumption that is not
generally valid. The classic signature of such events is missing transverse
momentum ($\Big/ \hspace{-0.3cm P_{T}}$) from the escaping photinos.
The analysis of the UA1 Collaboration gave special attention to this
characteristic \cite{Baer:1986au}.

Unfortunately, the data analyses of the Large Electron Positron
Collider at the CERN (LEP) were unable to follow the same approach. For
larger gaugino masses, it is unproductive to think in terms
of the photino, zino and neutrals higgsinos ($\tilde{h}_{1}^{0}$ and
$\tilde{h}_{2}^{0}$). Instead, one must consider the mixture of these
states giving four neutralinos $\tilde{\chi}^{0}_{j}$, $j=1,2,3,4$. In
a similar way the mixing of the charged gauginos with the charged
higgsinos gives two charginos $\tilde{\chi}^{\pm}_{i}$, $i=1,2$. The
dominant gluino decays then occur via $\tilde{g}\rightarrow
\bar{q}q\tilde{\chi}^{\pm}_{i}$ and $\tilde{g}\rightarrow
\bar{q}q\tilde{\chi}^{0}_{i}$
\cite{dress,tata,Baer:1986au,Baer:1990sc}.

The rate of the two-body decay $\tilde{g} \to \tilde{\gamma}g$ of the
gluinos was first analysed in Ref.~\cite{Haber:1983fc}, where this
decay rate was found to vanish for
$m_{\tilde{q}_{L}}=m_{\tilde{q}_{R}}$.  The partial width for the
gluino radiative decay was recomputed (for $m_{\tilde{\gamma}}=0$) in
Refs.~\onlinecite{Ma:1988ns,Barbieri:1987ed}. The most general result
for the radiative decay width of the gluinos was obtained in
Ref.~\onlinecite{Haber:1983fc} for the photino as the Lightest Supersymmetric
Particle (LSP)~\cite{Baer:1990sc}. Only with very massive gluinos
($m_{\tilde{g}}>(m_{q}+m_{\tilde{q}})$) do the two-body decays into
quark plus squark become kinematically acessible and rapidly dominate
the branching fraction \cite{dress,tata}.

Notwithstanding the appealing arguments favoring SUSY, no
supersymmetric particle has been found so far, in the first LHC runs
of up to 8~TeV CM energies \cite{resultssusysearches}. With the
increasing luminosity and energies up to 14 TeV in the next years,
however, the prospect for discoveries is still good.  The search is
aided by the ``Snowmass Points and Slopes'' (SPS) \cite{sps1}, a set
of benchmark points and parameter lines in the MSSM parameter space
corresponding to different scenarios in the quest for supersymmetry
(see Ref.~\onlinecite{{sps2}} for an instructive review).  The goal
here is to reconstruct the fundamental supersymmetric theory, and its
breaking mechanism, from the experimental data
\cite{sps1,sps2,sps}. Therefore, in various scenarios, given the SPS
convention, the neutralinos (photinos) are the lighter particles, while
the gluino is the most massive particle of the MSSM.  As indicated by
Table~(\ref{tab:tmasses}), each set of parameters leads to different
masses for the gluinos, squarks, photinos and selectrons, which are
the only relevant parameters in the study.

\begin{table}[htb]
\renewcommand{\arraystretch}{1.10}
\begin{center}
\normalsize
 \vspace{0.5cm}
\begin{tabular}{|c|c|c|c|c|}
\hline
\hline
Scenario & $m_{\tilde{g}}\, (GeV)$ & $M_{\tilde{q}}\, (GeV)$ & $m_{\tilde{\gamma}}\, (GeV)$ & $M_{\tilde{e}}\, (GeV)$ \\
\hline
\hline
 SPS1a & 595.2  & 539.9 & 96 & 202 \\
 SPS1b & 916.1  & 836.2 & 96 & 202 \\
 SPS2 & 784.4  & 1533.6 & 79 & 1456 \\
 SPS3 & 914.3  & 818.3 & 160 & 287 \\
 SPS4 & 721.0  & 732.2 & 118 & 448 \\
 SPS5 & 710.3  & 643.9 & 119 & 256 \\
 SPS6 & 708.5  & 641.3 & 189 & 264 \\
 SPS7 & 926.0  & 861.3 & 161 & 261 \\
 SPS8 & 820.5  & 1081.6 & 137 & 356 \\
 SPS9 & 1275.2  & 1219.2 & 175 & 319 \\
  \hline
\hline
\end{tabular}
\caption{Masses of gluinos, squarks, photinos and selectrons in the SPS scenarios.}
\label{tab:tmasses} 
\end{center}
\end{table}

Supersymmetric theories involve self-conjugate Majorana spinors. In
Sec.~\ref{sec:majorana} we review the prescription for writing Feynman
rules for Majorana particles, which are based on a well-defined
fermion flow, a procedure that is similar to the one leading to
Feynman amplitudes for Dirac fermions.
Secs.~\ref{sec:photinos}~and \ref{sec:gluinos}, which detail the
calculation of the differential cross section for the production of
photinos and gluinos, respectively, are followed by conclusions.

\section{Feynman Rules for Majorana Particles}
\label{sec:majorana}

Neutral particles may or may not have distinct antiparticles. While
Dirac fermions \cite{Dirac} have antiparticles, the
neutron being an example, the contrary is true for Majorana fermions
\cite{Majorana},  the field operators of which therefore satisfy the
equalities\footnote{Appendix D of Ref.~\onlinecite{kane} discusses
this subject in detail.} \cite{kane}
\begin{eqnarray}
\psi_{M}&=& \psi^{c}_{M}\equiv C \bar{\psi}^{T}_{M}, \nonumber \\
\bar{\psi}_{M}&=& \psi^{T}_{M}C,
\label{defmajoranaspinor}
\end{eqnarray}
where $\bar{\psi}\equiv \psi^{\dagger}\gamma^{0}$, while $C$ is the
charge conjugation matrix. The latter has the following properties
\begin{eqnarray}
C^{\dagger}&=&C^{-1}, \,\
C^{T}=-C, \nonumber \\
C^{-1}\Gamma_{i}C&=&\Gamma^{T}_{i}, \,\ \mbox{for} \,\ \Gamma_{i}=I_{4 \times 4}, \imath \gamma_{5}, \gamma_{m}\gamma_{5}, \nonumber \\ 
C^{-1}\Gamma_{i}C&=&- \Gamma^{T}_{i}, \,\ \mbox{for} \,\
\Gamma_{i}= \gamma_{m}, \sigma_{m,n}= \frac{\imath}{2}[ \gamma_{m}, \gamma_{n}] .
\label{definicaoc}
\end{eqnarray}

The $\Gamma_{i}$ have been chosen such that
\begin{equation}
\Gamma^{\dagger}_{i}= \eta_{i}    \gamma^{0}\Gamma_{i}\gamma^{0},
\end{equation}
with no summation over $i$, where $\eta_{i}$ is defined as
\begin{eqnarray}
\eta_{i}= \left\{ \begin{array}{ccc}
1 &\mbox{for} \,\ i= &I_{4 \times 4},i \gamma_{5}, \gamma_{m},  \\
-1 &\mbox{for}\,\ i= & \gamma_{m}, \sigma_{m,n}\,.
\end{array}
\right. 
\label{constraintatcouplingconstant1}
\end{eqnarray}
%We now note that
%\begin{eqnarray}
%C^{-1}\gamma_{m}C&=&- \gamma^{T}_{m} \rightarrow
%\gamma_{m}C=-C \gamma^{T}_{m}.
%\end{eqnarray}

In general, the $u$ and $v$ spinors for either Dirac or Majorana
fermions are related by the equalities
\begin{eqnarray}
u^{c}\equiv C \bar{u}^{T}=v, \,\ \bar{u}^{(s)T}=C^{-1}v^{(s)}, \,\ 
v^{(s)T}=\bar{u}^{(s)}C^{T}, \nonumber \\
v^{c}\equiv C \bar{v}^{T}=u, \,\ \bar{v}^{(s)T}=C^{-1}u^{(s)}, \,\ 
u^{(s)T}=\bar{v}^{(s)}C^{T},
\label{novasrelacoes}
\end{eqnarray}
where $s= \pm1/2$ labels the spin. 

The Feynman rules for Majorana fermions, by contrast with those for
Dirac fermions, involve vertices and propagators with clashing
arrows. As a consequence, charge-conjugation
matrices appear in the Feynman rules for vertices and propagators, as
discussed in Refs.~\onlinecite{dress,tata,ait1,kane}.

\subsection{Problems in defining Feynman rules for a Majorana Field.}

In the Standard Model (SM) \cite{sg} all the interactions conserve
both the Baryon number ($B$) and the Lepton number ($L$). By contrast,
the MSSM \cite{R,ssm,grav,kane} comprises interactions that
violate the conservation of fermion number, because the Majorana
Fermions lack distinct antiparticles. Their self-conjugacy allows
for a variety of different contractions, which acquire different signs
due to the anticommutation of fermionic operators
\cite{kane,denner,denner1}.

The usual Dirac field spinor expansion is given by the expression \cite{tata}
\begin{equation}
\Psi_{D}(x)= \int \frac{d^{3}k}{(2 \pi)^{3}} \frac{1}{2E_{\vec{k}}}\sum_{s= \mp 1/2} \left[
c_{\vec{k},s}u^{(s)}(k)e^{-ikx}+ d^{\dagger}_{\vec{k},s}v^{(s)}(k)e^{ikx} \right] ,
\end{equation}
where $c(c^{\dagger})$ and $d(d^{\dagger})$ are annihilation
(creation) operators satisfying
\begin{eqnarray}
\left[ c_{\vec{k},s},c^{\dagger}_{\vec{l},r} \right]&=&(2 \pi)^{3}2E_{\vec{k}}\delta_{sr}\delta^{3} \left( \vec{k}- \vec{l} \right), \,\
\left[ c_{\vec{k},s},c_{\vec{l},r} \right]= \left[ c^{\dagger}_{\vec{k},s},c^{\dagger}_{\vec{l},r} \right]=0, \nonumber \\
\left[ d_{\vec{k},s},d^{\dagger}_{\vec{l},r} \right]&=&(2 \pi)^{3}2E_{\vec{k}}\delta_{sr}\delta^{3} \left( \vec{k}- \vec{l} \right), \,\
\left[ d_{\vec{k},s},d_{\vec{l},r} \right]= \left[ d^{\dagger}_{\vec{k},s},d^{\dagger}_{\vec{l},r} \right]=0.
\end{eqnarray}

To quantize the Dirac spinor field $\Psi_{D}$ one requires that
\begin{eqnarray}
\{ \Psi_{Da}(x), \Psi^{\dagger}_{Db}(y) \} &=& \delta_{ab}\delta^{3}( \vec{x}- \vec{y}), \nonumber \\
\{ \Psi_{Da}(x), \Psi_{Db}(y) \}&=& \{ \Psi^{\dagger}_{Da}(x), \Psi^{\dagger}_{Db}(y) \}=0.
\end{eqnarray}

For a Dirac spinor, therefore, we can write the following expressions:
\begin{eqnarray}
\langle 0|T \left\{ \Psi_{Da}(x) \bar{\Psi}_{Db}(y) \right\} | 0 \rangle &=&S_{Fab}(x-y), \nonumber \\
\langle 0|T \left\{ \Psi_{Da}(x) \Psi_{Db}(y) \right\} | 0 \rangle &=& \langle 0|T \left\{ \bar{\Psi}_{Da}(x) \bar{\Psi}_{Db}(y) \right\} | 0 \rangle =0.
\end{eqnarray}

We represent a fermion in a Feynman diagram by a solid line. For a Dirac fermion 
each line carries an arrow indicating the fermion number flow.

The analogous expressions for the Majorana fermions \cite{tata} are
\begin{eqnarray}
\Psi_{M}(x)&=& \int \frac{d^{3}k}{(2 \pi)^{3}} \frac{1}{2E_{\vec{k}}}\sum_{s= \mp 1/2} \left[
c_{\vec{k},s}u^{(s)}(k)e^{-ikx}+ c^{\dagger}_{\vec{k},s}v^{(s)}(k)e^{ikx} \right] , \nonumber \\
\left[ c_{\vec{k},s},c^{\dagger}_{\vec{l},r} \right]&=&(2 \pi)^{3}2E_{\vec{k}}\delta_{sr}\delta^{3} \left( \vec{k}- \vec{l} \right), \,\
\left[ c_{\vec{k},s},c_{\vec{l},r} \right]= \left[ c^{\dagger}_{\vec{k},s},c^{\dagger}_{\vec{l},r} \right]=0, \nonumber \\
\langle 0|T \left\{ \Psi_{Ma}(x) \bar{\Psi}_{Mb}(y) \right\} | 0 \rangle &=&S_{Fab}(x-y), \nonumber \\
\langle 0|T \left\{ \Psi_{Ma}(x) \Psi_{Mb}(y) \right\} | 0 \rangle &=& \langle 0|T \left\{ \Psi_{Ma}(x) \bar{\Psi}_{Mc}(y) \right\} | 0 \rangle C^{T}_{cb}=
S_{Fac}(x-y)C^{T}_{cb}, \nonumber \\
\langle 0|T \left\{ \bar{\Psi}_{Ma}(x) \bar{\Psi}_{Mb}(y) \right\} | 0 \rangle &=& C^{T}_{ac}\langle 0|T \left\{ \Psi_{Mc}(x) \bar{\Psi}_{Mb}(y) \right\} | 0 \rangle
=C^{T}_{ac}S_{Fcb}(x-y), \nonumber \\
\label{propagatormajoranatres}
\end{eqnarray}
where $c$ and $c^{\dagger}$ are annihilation and creation
operators,\footnote{The condition $c_{\vec{k},s}=d_{\vec{k},s}$
  implies the identity of the particle and antiparticle quanta of this
  field.}  and we have to include these contractions to compute
matrix elements of operators involving products of Majorana spinor
fields \cite{tata,kane}. By contrast with Dirac lines, Majorana lines
carry no arrows.

For Dirac fields the internal propagator reads
\begin{equation}
\langle 0|T \left\{ \Psi(x) \bar{\Psi}(y) \right\} | 0 \rangle \rightarrow 
\frac{1}{\not\hspace{-.4ex}{P} -m}=S(P),
\label{eq:intprop}
\end{equation}
where the propagating fields $\Psi$ carries a momentum $P$. 

The contraction with the external operators is given by the expressions
\cite{denner,denner1}
\begin{eqnarray}
\langle 0|T \left\{ \Psi(x) c^{\dagger}(p_{i},s_{i}) \right\} | 0 \rangle &\rightarrow& u(p_{i},s_{i}), \nonumber \\
\langle 0|T \left\{ c(p_{i},s_{i}) \bar{\Psi}(x) \right\} | 0 \rangle &\rightarrow& \bar{u}(p_{i},s_{i}), \nonumber \\
\langle 0|T \left\{ \Psi(x) d^{\dagger}(p_{i},s_{i}) \right\} | 0 \rangle &\rightarrow& \bar{v}(p_{i},s_{i}), \nonumber \\
\langle 0|T \left\{ d(p_{i},s_{i}) \bar{\Psi}(x) \right\} | 0 \rangle &\rightarrow& v(p_{i},s_{i}).
\label{eq:extprop}
\end{eqnarray}

\begin{figure}[ht]
\begin{center}
\vglue -0.009cm %\includegraphics{assi1f1.eps}
\mbox{\epsfig{file=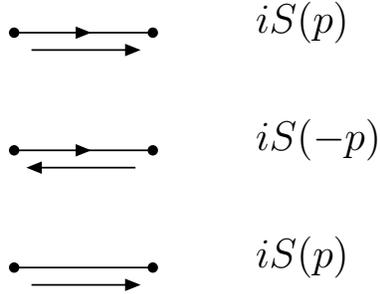,width=0.3\textwidth,angle=0}}
\end{center}
\caption{Feynman rules for an internal Dirac fermion line. For each diagram, the upper line represents the fermion momentum, and the lower one represents the fermion flow.}
\label{linhainternapropdirac}
\end{figure}

\begin{figure}[ht]
\begin{center}
\vglue -0.009cm %\includegraphics{assi1f1.eps}
\mbox{\epsfig{file=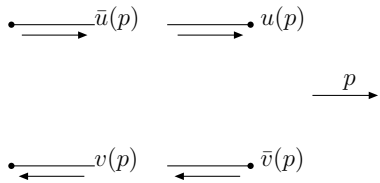,width=0.3\textwidth,angle=0}}
\end{center}
\caption{Feynman rules for an external Dirac fermion line. The momentum $p$ flows from the left to the right.}
\label{linhaexternapropdirac}
\end{figure}

\begin{figure}[ht]
\begin{center}
\vglue -0.009cm %\includegraphics{assi1f1.eps}
\mbox{\epsfig{file=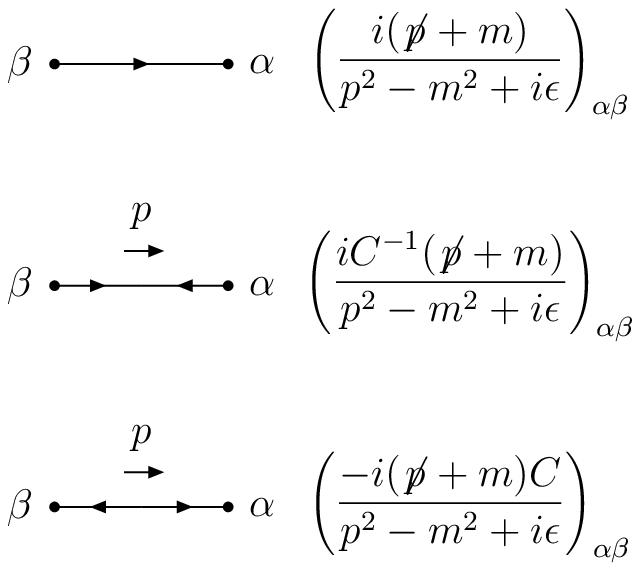,width=0.3\textwidth,angle=0}}
\end{center}
\caption{Feynmann rules for Majorana fermions propagators with orientation (thin arrow). The momentum $p$ flows from left to right.}
\label{linhainternapropmajorana}
\end{figure}

In general, each Dirac field is associated with the usual propagator
$S(P)$, see Eq.~(\ref{eq:intprop}), and the ``reversed" one $S(-P)$,
as drawn in Fig.~\ref{linhainternapropdirac}, as well as with the
usual spinors and their ``reversed" counterparts, given by
Eq.~(\ref{eq:extprop}), which are depicted in Fig.~\ref{linhaexternapropdirac}. For
Majorana fermions, since arrows cannot be drawn to indicate the
fermion-number flow, we only have the usual propagator $S(P)$ and
spinors, not the reversed ones, as shown by
Fig.~\ref{linhainternapropmajorana}. Notice that the propagator has
clashing arrows \cite{denner,denner1}.

The most generic lagrangian ${\cal L}$ for Majorana fields $\lambda$ and
Dirac fields $\psi$ \cite{kane}, augmented by a pure Dirac interaction
term \cite{denner,denner1}, can be expressed by the following expression:
\begin{eqnarray}
{\cal L}&=& \frac{1}{2} \bar{\lambda}_{a}\left( \imath \gamma^{m}\partial_{m}-M_{a}  
\right) \lambda_{a}+ \bar{\psi}_{a}\left( \imath \gamma^{m}\partial_{m}-m_{a}  
\right) \psi_{a}+ \frac{1}{2}g^{i}_{abc}\bar{\lambda}_{a}\Gamma_{i}\lambda_{b}\Phi_{c} \nonumber \\
&+& \frac{1}{2}g^{i*}_{abc}\bar{\lambda}_{b}\Gamma_{i}\lambda_{a}\Phi^{*}_{c}+
\kappa^{i}_{abc}\bar{\lambda}_{a}\Gamma_{i}\psi_{b}\Phi^{*}_{c}+
\kappa^{i*}_{abc}\bar{\psi}_{b}\Gamma_{i}\lambda_{a}\Phi_{c}+
h^{i}_{abc}\bar{\psi}_{a}\Gamma_{i}\psi_{b}\Phi_{c}, \nonumber \\
\label{fmdv}
\end{eqnarray}
where $\Gamma_{i}$ is defined in Eq.~(\ref{definicaoc}),
and $g^{i}_{abv}, \kappa^{i}_{abc}$ and $h^{i}_{abc}$ are coupling constants. 

The field $\Phi$ summarizes scalar and vector fields. Using
Eq.~(\ref{definicaoc}) and the fact that fermion fields anticommute in
the third term on the right-hand side, we find that the following constraint
must be satisfied
\begin{equation}
g^{i}_{abc}= \eta_{i} g^{i}_{bac}\,,
\label{constraintatcouplingconstant}
\end{equation}
with no summation over $i$, and $\eta$ given by
Eq.~(\ref{constraintatcouplingconstant1}).

The second problem, as shown in Ref.~\cite{kane}, is the
following. The fourth term on the right-hand side of Eq.~(\ref{fmdv})
can be rewritten as
\begin{eqnarray}
g^{i}_{abc}\bar{\lambda}_{b}\Gamma_{i}\lambda_{a}\Phi^{*}_{c}&=&-g^{i}_{abc}\bar{\lambda}^{T}_{a}\left( C^{-1}\Gamma_{i} \right)\bar{\lambda}_{b}\Phi^{*}_{c},
\nonumber \\
g^{i}_{abc}\bar{\lambda}_{b}\Gamma_{i}\lambda_{a}\Phi^{*}_{c}&=&g^{i}_{abc}\bar{\lambda}_{a}\left( \Gamma_{i}C \right)\bar{\lambda}^{T}_{b}\Phi^{*}_{c},
\end{eqnarray}
and hence the Feynman rules for this term may then be given by any of
the following conventions
\begin{eqnarray}
&&i g^{i}_{abc} \Gamma = -ig^{i}_{abc} \left( C^{-1}\Gamma_{i} \right), \nonumber \\
&&i g^{i}_{abc} \Gamma = ig^{i}_{abc} \left( \Gamma C \right),
\end{eqnarray}
which seem to give evidence of sign ambiguity. 

Another problem derives from the location of the $C$ operator
\cite{kane}. The self-conjugacy allows for a variety of different
contractions, which acquire different signs originating from the
anticommutation of fermionic operators
\cite{kane,denner,denner1}. In this approach the relative sign of
interfering Feynman graphs cannot be read off the graphs, but has to
be determined independently from the Wick contractions. This method is
unwieldy in such practical calculations as the production  of photinos and gluinos.

\begin{figure}[t]
\begin{center}
\vglue -0.009cm %\includegraphics{assi1f1.eps}
\mbox{\epsfig{file=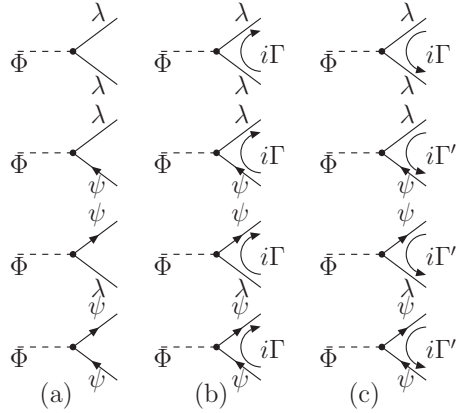,width=0.36\textwidth,angle=0}}
\end{center}
\caption{Feynmann rules for the interaction (from Eq.~(\ref{fmdv})) of a scalar field with one Majorana ($\lambda$)
fermion. From Refs.~\onlinecite{denner,denner1}.}
\label{majint}
\end{figure}

There is however an alternative way to define Feynman rules for
Majorana fermions. Since the fermion-flow rule is violated, we may
introduce a continuous fermion-flow orientation for each fermion line,
as in Refs.~\onlinecite{gates,denner,denner1}.\footnote{We only need the
  familiar Dirac propagator and only vertices without explicit
  charge-conjugation matrices.} This forces us to introduce two
analytical expressions for each vertex, one for fermion flow parallel
and the other for fermion flow antiparallel to the flow of the fermion
number. Therefore, for Majorana fermions, only the usual
spinors are present, as in Fig.~\ref{majint}.

The Feynman amplitudes can be obtained from the following procedure \cite{denner,denner1}:
\begin{enumerate}
\item Draw all possible Feynman diagrams for a given process;
\item Fix an arbitrary orientation (fermion flow) for each fermion chain;
\item Start at an external leg (for closed loops at some arbitrary propagator) and write down the Dirac 
matrices proceeding opposite to the chosen orientation (fermion flow) through the chain in agreement with 
Fig.~\ref{linhaexternapropdirac};
\item Apply the corresponding analytical expressions;
\item Multiply by a factor $(-1)$ for every closed loop;
\item Multiply by the permutation parity of the spinors in the obtained analytical expression with the respect to some reference order;
\item As far as the determination of the combinatorial factor is concerned, Majorana fermions behave
exactly like real scalar or vector fields. 
\end{enumerate}

We can understand the last item in the following way: for Majorana fermions there 
are two equivalent non-vanishing Wick contractions, i.~e.,
\begin{eqnarray}
\bar{\chi}\Gamma \chi = \overline{\tilde{\chi}}\Gamma^{\prime}\tilde{\chi}.
\end{eqnarray}
For Majorana fermions $\tilde{\chi}= \chi$, an equality that, together with
Eq.~(\ref{constraintatcouplingconstant}), allows us to show that
\begin{eqnarray}
\Gamma^{\prime}= \Gamma.
\end{eqnarray}

These two contractions yield the same result and cancel the factor $1/2$ in 
the corresponding interaction term. This is 
exactly what happens for real scalar and vectors fields. %, with this we justify the last item above. 
The analytical expressions are independent of the chosen orientation,
i.~e., of the fermion flow, as shown by Refs.~\onlinecite{denner,denner1}.

This set of rules, which will be shown in the next sections to
simplify pratical calculations, can be combined with the FeynArts
program \cite{Hahn:2000kx} to calculate the differential cross
sections for photino and gluino production.

In general, to calculate the square amplitudes involving Dirac
fermions, we use the following projection operators in diagrams
involving Majorana fermions, which are obtained from Eq.~(\ref{novasrelacoes}):
\begin{eqnarray}
\sum_{s}u^{(s)}(P)v^{(s)T}(P)&=&\left( \sum_{s}u^{(s)}(P)\bar{u}^{(s)}(P) \right)C^{T}=( \not\hspace{-.4ex}{P}+M)C^{T}, \nonumber \\
\sum_{s}v^{(s)}(P)u^{(s)T}(P)&=&\left( \sum_{s}v^{(s)}(P)\bar{v}^{(s)}(P) \right)C^{T}=( \not\hspace{-.4ex}{P}-M)C^{T}, \nonumber \\
\sum_{s}\bar{u}^{(s)T}(P)\bar{v}^{(s)}(P)&=&C^{-1}\left( \sum_{s}v^{(s)}(P)\bar{v}^{(s)}(P) \right)=C^{-1}( \not\hspace{-.4ex}{P}-M), \nonumber \\
\sum_{s}\bar{v}^{(s)T}(P)\bar{u}^{(s)}(P)&=&C^{-1}\left( \sum_{s}u^{(s)}(P)\bar{u}^{(s)}(P) \right)=C^{-1}( \not\hspace{-.4ex}{P}+M), \nonumber \\
\label{projecaomajorana}
\end{eqnarray}
where $C$ is the charge conjugation matrix satisfying Eq.~(\ref{definicaoc}).

\section{Photino Production}
\label{sec:photinos}

\begin{figure}[ht]
\begin{center}
\vglue -0.009cm %\includegraphics{assi1f1.eps}
\mbox{\epsfig{file=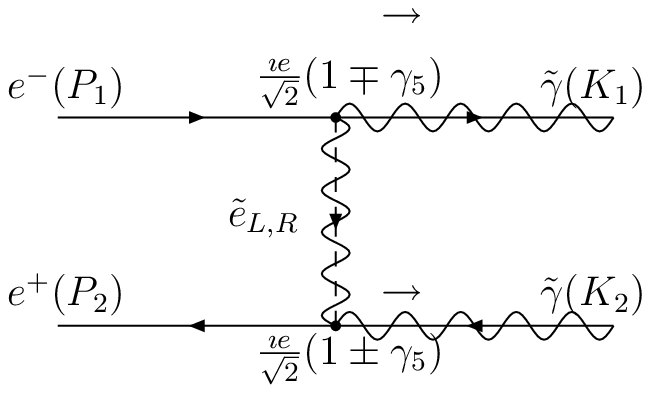,width=0.5\textwidth,angle=0}} 
\vglue -0.009cm %\includegraphics{assi1f1.eps}
\mbox{\epsfig{file=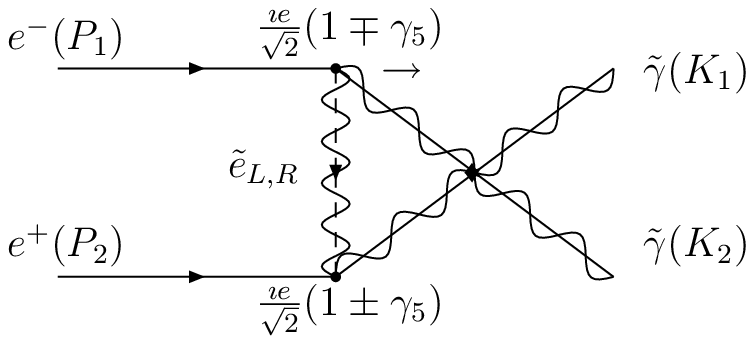,width=0.5\textwidth,angle=0}} 
\end{center}
\caption{Feynman diagram of the process $e^{-}e^{+}\rightarrow
  \tilde{\gamma}\tilde{\gamma}$. The electron fixes the orientation
  (fermion flow) for each lepton. The positron line has therefore the
  opposite direction, and a continuous line results in the diagram.}
\label{cap2.3}
\end{figure}

This section shows how to calculate the differential cross section of
the $e^{-}e^{+} \to \tilde{\gamma}\tilde{\gamma}$ process, where
$\tilde{\gamma}$ is the photino field in the SQED context. This
process, which conserves the lepton number, takes place via
$t$-channel $\tilde{e}^{-}_{L}$- and $\tilde{e}^{-}_{R}$-exchange (see
Fig.~\ref{cap2.3}) as shown by Fayet~\cite{Fayet:1982ky}.  We will
assume that $\tilde{e}^{-}_{L}$- and $\tilde{e}^{-}_{R}$ are mass
eigenstates, because we are in the domain of SQED. Both $e$ and
$\tilde{e}$ carry one unit of lepton number.  The $t$- and $u$-channel
exchanges correspond to the cases where the fermion lines are
uncrossed and crossed, respectively.

Today we know that in the context of the MSSM the photino is a
gaugino and mixes with the neutral higgsinos to yield the
neutralinos as the mass eigenstates
\cite{dress,tata,Bartl:1989ms}. Neutralino pair production in
$e^{-}e^{+}$ collisions was first studied in \cite{Carena:1986jp},
where it was shown that this production takes place via the
$s$-channel $Z$-exchange and $t$-channel $\tilde{e}^{-}_{L}$- and
$\tilde{e}^{-}_{R}$-exchange. On the other hand,
the LSP in some minimal Supergravity (mSUGRA) scenarios can be a light photino,
i.~e., $\tilde{\chi}^{0}_{1} \approx \tilde{\gamma}$
\cite{sps1,sps2,sps}, with an acceptable cosmological abundance
\cite{Olive:1994qq}.

In $e^{-}e^{+}$ collisions, photinos are produced in the following 
reaction:
\begin{equation}
e^{-}(P_{1})+e^{+}(P_{2}) \rightarrow \tilde{\gamma}(K_{1})+ \tilde{\gamma}(K_{2})
\end{equation}
as shown in Fig.~\ref{cap2.3}, where $\tilde{\gamma}$ is the photino, and the particle four-momenta are specified in parentheses. 
As already explained, the photino is its own antiparticle.

The amplitudes for $e^{-}(P_{1})e^{+}(P_{2})\rightarrow \tilde{\gamma}(K_{1})\tilde{\gamma}(K_{2})$ are
\begin{eqnarray}
{\cal M}_{a}&=&- \frac{e^{2}}{2}
\bar{u}(K_{1})(1- \gamma_{5})u(P_{1})
\frac{1}{t-M^{2}_{\tilde{e}_{L}}}
\bar{v}(P_{2})(1+ \gamma_{5})v(K_{2}), \nonumber \\
{\cal M}_{b}&=& \frac{e^{2}}{2}
\bar{u}(K_{1})(1+ \gamma_{5})u(P_{1})
\frac{1}{t-M^{2}_{\tilde{e}_{R}}}
\bar{v}(P_{2})(1- \gamma_{5})v(K_{2}), \nonumber \\
{\cal M}_{c}&=&- \frac{e^{2}}{2}
\bar{u}(K_{2})(1- \gamma_{5})u(P_{1})
\frac{1}{u-M^{2}_{\tilde{e}_{L}}}
\bar{v}(P_{2})(1+ \gamma_{5})v(K_{1}), \nonumber \\
{\cal M}_{d}&=&\frac{e^{2}}{2}
\bar{u}(K_{2})(1+ \gamma_{5})u(P_{1})
\frac{1}{u-M^{2}_{\tilde{e}_{R}}}
\bar{v}(P_{2})(1- \gamma_{5})v(K_{1}),
\label{amp}
\end{eqnarray}
where $s,t,u$ are the Mandelstam variables, defined as 
\begin{eqnarray}
s&=&(P_{1}+P_{2})^{2}=(K_{1}+K_{2})^{2}, \nonumber \\
t&=&(P_{1}-K_{1})^{2}=(P_{2}-K_{2})^{2}, \nonumber \\
u&=&(P_{1}-K_{2})^{2}=(P_{2}-K_{1})^{2}, \nonumber \\
s+t+u&=&2m^{2}_{e}+2m^{2}_{\tilde{\gamma}}.
\label{mandelstamvar}
\end{eqnarray}
Here $m_{e}$ and $m_{\tilde{\gamma}}$ are the electron and the photino 
masses, respectively.

The next step is to calculate $|{\cal M}_{a}+ {\cal M}_{b}- {\cal
  M}_{c}- {\cal M}_{d}|^{2}$, where the relative negative signs are
due to the Pauli statistics. Summing over initial and final spins and using the 
usual projection operator over the positive and negative energy
states, we obtain (see Ref.~\onlinecite{danusa} for more details on the
algebric manipulations) the following expressions:
\begin{eqnarray}
|{\cal M}_{a}|^{2}&=&\frac{16e^{4}}{(t-M^{2}_{\tilde{e}_{L}})^{2}}(P_{1} \cdot K_{1})(P_{2} \cdot K_{2})
= \frac{4e^{4}}{(t-M^{2}_{\tilde{e}_{L}})^{2}}\left( t-m^{2}_{e}-m_{\tilde{\gamma}}^{2} \right)^{2},   \nonumber \\
{\cal M}^{\dagger}_{a}{\cal M}_{b}&=&{\cal M}_{b}{\cal M}^{\dagger}_{a}= 
\frac{e^{4}}{4(t-M^{2}_{\tilde{e}_{L}})(t-M^{2}_{\tilde{e}_{R}})}(8m_{e}m_{\tilde{\gamma}})^{2},  \nonumber \\
|{\cal M}_{b}|^{2}&=&\frac{16e^{4}}{(t-M^{2}_{\tilde{e}_{R}})^{2}}(P_{1} \cdot K_{1})(P_{2} \cdot K_{2})
= \frac{4e^{4}}{(t-M^{2}_{\tilde{e}_{R}})^{2}}\left( t-m^{2}_{e}-m_{\tilde{\gamma}}^{2} \right)^{2}, \nonumber \\
|{\cal M}_{c}|^{2}&=&\frac{16e^{4}}{(u-M^{2}_{\tilde{e}_{L}})^{2}}(P_{1} \cdot K_{2})(P_{2} \cdot K_{1})
=\frac{4e^{4}}{(u-M^{2}_{\tilde{e}_{L}})^{2}} (u-m^{2}_{e}-m^{2}_{\tilde{\gamma}})^{2},  \nonumber \\
{\cal M}^{\dagger}_{c}{\cal M}_{d}&=&{\cal M}^{\dagger}_{d}{\cal M}_{c}=
\frac{e^{4}}{4(u-M^{2}_{\tilde{e}_{R}})(u-M^{2}_{\tilde{e}_{L}})}(8m_{e}m_{\tilde{\gamma}})^{2}, \nonumber \\
|{\cal M}_{d}|^{2}&=&\frac{16e^{4}}{(u-M^{2}_{\tilde{e}_{R}})^{2}}(P_{1} \cdot K_{2})(P_{2} \cdot K_{1})
= \frac{4e^{4}}{(u-M^{2}_{\tilde{e}_{R}})^{2}}(u-m^{2}_{e}-m^{2}_{\tilde{\gamma}})^{2}. \nonumber \\  
\end{eqnarray}

Lets work out the interference terms in detail. On this case we need
again to sum over initial and final spins. From
Eq.~(\ref{projecaomajorana}), we find that
\begin{eqnarray}
{\cal M}^{\dagger}_{a}{\cal M}_{c}&=&{\cal M}^{\dagger}_{c}{\cal M}_{a}= 
\frac{e^{4}}{4(u-M^{2}_{\tilde{e}_{L}})(t-M^{2}_{\tilde{e}_{L}})} 
\left\{ \mbox{Tr}\left[ (1- \gamma_{5}) ( \not\hspace{-.4ex}{P}_{1}\hspace{.1ex}+m_{e})(1+ \gamma_{5}) \right. \right. \nonumber \\
&\cdot& \left. \left. 
(\not\hspace{-.4ex}{K}_{1}+m_{\tilde{\gamma}}) C^{T}(1+ \gamma_{5})^{T}( \not\hspace{-.4ex}{P}_{2}-m_{e})^{T}
(1- \gamma_{5})^{T}C^{-1}(\not\hspace{-.4ex}{K}_{2}+m_{\tilde{\gamma}}) \right] \right\},
\nonumber \\
{\cal M}^{\dagger}_{a}{\cal M}_{d}&=&{\cal M}^{\dagger}_{d}{\cal M}_{a}= 
\frac{e^{4}}{4(t-M^{2}_{\tilde{e}_{L}})(u-M^{2}_{\tilde{e}_{R}})} 
\left\{ \mbox{Tr}\left[ (1+ \gamma_{5}) ( \not\hspace{-.4ex}{P}_{1}\hspace{.1ex}+m_{e})(1+ \gamma_{5}) \right. \right. \nonumber \\
&\cdot& \left. \left.
(\not\hspace{-.4ex}{K}_{1}+m_{\tilde{\gamma}}) C^{T}(1- \gamma_{5})^{T}( \not\hspace{-.4ex}{P}_{2}-m_{e})^{T}
(1+ \gamma_{5})^{T}C^{-1}(\not\hspace{-.4ex}{K}_{2}+m_{\tilde{\gamma}}) \right] \right\},
\nonumber \\
{\cal M}^{\dagger}_{b}{\cal M}_{c}&=&{\cal M}^{\dagger}_{c}{\cal M}_{b}= 
\frac{e^{4}}{4(t-M^{2}_{\tilde{e}_{L}})(u-M^{2}_{\tilde{e}_{R}})} 
\left\{ \mbox{Tr}\left[ (1- \gamma_{5}) ( \not\hspace{-.4ex}{P}_{1}\hspace{.1ex}+m_{e})(1+ \gamma_{5}) \right. \right. \nonumber \\
&\cdot& \left. \left. 
(\not\hspace{-.4ex}{K}_{1}+m_{\tilde{\gamma}}) C^{T}(1- \gamma_{5})^{T}( \not\hspace{-.4ex}{P}_{2}-m_{e})^{T}
(1- \gamma_{5})^{T}C^{-1}(\not\hspace{-.4ex}{K}_{2}+m_{\tilde{\gamma}}) \right] \right\},
\nonumber \\
{\cal M}^{\dagger}_{b}{\cal M}_{d}&=&{\cal M}^{\dagger}_{d}{\cal M}_{b}= 
\frac{e^{4}}{4(u-M^{2}_{\tilde{e}_{R}})(t-M^{2}_{\tilde{e}_{R}})} 
\left\{ \mbox{Tr}\left[ (1+ \gamma_{5}) ( \not\hspace{-.4ex}{P}_{1}\hspace{.1ex}+m_{e})(1- \gamma_{5}) \right. \right. \nonumber \\
&\cdot& \left. \left. 
(\not\hspace{-.4ex}{K}_{1}+m_{\tilde{\gamma}}) C^{T}(1- \gamma_{5})^{T}( \not\hspace{-.4ex}{P}_{2}-m_{e})^{T}
(1+ \gamma_{5})^{T}C^{-1}(\not\hspace{-.4ex}{K}_{2}+m_{\tilde{\gamma}}) \right] \right\}\,.
\nonumber \\
\end{eqnarray}

We now need to calculate the quantity
\begin{eqnarray}
C^{T}(1+ \gamma_{5})^{T}( \not\hspace{-.4ex}{P}_{2}-m_{e})^{T}
(1- \gamma_{5})^{T}C^{-1}. 
\label{eq28}
\end{eqnarray}

Notice taken that $\left[ C, \gamma_{5} \right] =0$ and 
$\gamma^{T}_{5}= \gamma_{5}$, with the first two equalities in
Eq.~(\ref{definicaoc})  we can show that
\begin{eqnarray}
C^{T}(1+ \gamma_{5})^{T}&=&(1^{T}- \gamma^{T}_{5})C^{T}=-(1- \gamma_{5})C, 
\nonumber \\
(1- \gamma_{5})^{T}C^{-1}&=&C^{T}(1^{T}+ \gamma^{T}_{5})=C^{-1}(1+ \gamma_{5}), 
\end{eqnarray}
and using the last equality in Eq.~(\ref{definicaoc}), we can rewrite
Eq.~(\ref{eq28}) in the form
\begin{eqnarray}
&-&(1- \gamma_{5})(P^{m}_{2}C \gamma^{T}_{m}C^{-1}-m_{e}CC^{-1})(1+ \gamma_{5})=
-(1- \gamma_{5})(-P^{m}_{2}\gamma_{m}-m_{e})(1+ \gamma_{5}) \nonumber \\
&=&(1- \gamma_{5})( \not\hspace{-.4ex}{P}_{2}+m_{e})(1+ \gamma_{5}).
\end{eqnarray}

We have therefore shown the following relation to hold:
\begin{eqnarray}
C^{T}(1+ \gamma_{5})^{T}( \not\hspace{-.4ex}{P}_{2}-m_{e})^{T}(1- \gamma_{5})^{T}C^{-1}&=&
(1- \gamma_{5})( \not\hspace{-.4ex}{P}_{2}+m_{e})(1+ \gamma_{5}). \nonumber \\
\end{eqnarray}

Similarly, we can show that
\begin{eqnarray}
C^{T}(1- \gamma_{5})^{T}( \not\hspace{-.4ex}{P}_{2}-m_{e})^{T}(1+ \gamma_{5})^{T}C^{-1}&=&
(1+ \gamma_{5})( \not\hspace{-.4ex}{P}_{2}+m_{e})(1- \gamma_{5}). \nonumber \\
\label{cprop}
\end{eqnarray}

Next, using the trace techniques we find that
\begin{eqnarray}
{\cal M}^{\dagger}_{a}{\cal M}_{c}&=&{\cal M}^{\dagger}_{c}{\cal M}_{a}= 
\frac{8e^{4}m^{2}_{\tilde{\gamma}}}{(u-M^{2}_{\tilde{e}_{L}})(t-M^{2}_{\tilde{e}_{L}})} (P_{1} \cdot P_{2}) \nonumber \\
&=&\frac{8e^{4}m^{2}_{\tilde{\gamma}}}{(u-M^{2}_{\tilde{e}_{L}})(t-M^{2}_{\tilde{e}_{L}})} 
\left( \frac{s}{2}-m^{2}_{e} \right), \nonumber \\
{\cal M}^{\dagger}_{a}{\cal M}_{d}&=&{\cal M}^{\dagger}_{d}{\cal M}_{a}= 
\frac{e^{4}}{4(t-M^{2}_{\tilde{e}_{L}})(u-M^{2}_{\tilde{e}_{R}})}(8m_{e}m_{\tilde{\gamma}})^{2}, \nonumber \\
{\cal M}^{\dagger}_{b}{\cal M}_{c}&=&{\cal M}^{\dagger}_{c}{\cal M}_{b}= 
\frac{e^{4}}{4(t-M^{2}_{\tilde{e}_{R}})(u-M^{2}_{\tilde{e}_{L}})}(8m_{e}m_{\tilde{\gamma}})^{2},  \nonumber \\
{\cal M}^{\dagger}_{b}{\cal M}_{d}&=&{\cal M}^{\dagger}_{d}{\cal M}_{b}= 
\frac{8e^{4}m^{2}_{\tilde{\gamma}}}{(u-M^{2}_{\tilde{e}_{R}})(t-M^{2}_{\tilde{e}_{R}})}(P_{1} \cdot P_{2}) \nonumber \\
&=&\frac{8e^{4}m^{2}_{\tilde{\gamma}}}{(u-M^{2}_{\tilde{e}_{R}})(t-M^{2}_{\tilde{e}_{R}})} 
\left( \frac{s}{2}-m^{2}_{e} \right)\,. \nonumber \\
\label{pinterm}
\end{eqnarray}

The differential cross section in the %$M^{2}_{\tilde{e}_{L}}=M^{2}_{\tilde{e}_{R}}=M^{2}_{\tilde{e}}$ 
$M_{\tilde{e}_{L}}=M_{\tilde{e}_{R}}=M_{\tilde{e}}$ limit is then
given by the relation
\begin{eqnarray}
\frac{d \sigma}{d \Omega}(e^{-}e^{+}\rightarrow \tilde{\gamma}\tilde{\gamma})
&=&\frac{\alpha^{2}}{4s} \sqrt{\frac{s-4m^{2}_{\tilde{\gamma}}}{s-4m^{2}_{e}}} \left[
\left( \frac{t-m^{2}_{\tilde{\gamma}}-m^{2}_{e}}{t-M^{2}_{\tilde{e}}} \right)^{2}+ 
\left( \frac{u-m^{2}_{\tilde{\gamma}}-m^{2}_{e}}{u-M^{2}_{\tilde{e}}} \right)^{2} \right. \nonumber \\
&+& \left. 
\left( \frac{2m_{e}m_{\tilde{\gamma}}}{t-M^{2}_{\tilde{e}}} \right)^{2}+
\left( \frac{2m_{e}m_{\tilde{\gamma}}}{u-M^{2}_{\tilde{e}}} \right)^{2}+
\left(
\frac{16m^{2}_{e}m^{2}_{\tilde{\gamma}}-2sm^{2}_{\tilde{\gamma}}}{(t-M^{2}_{\tilde{e}})(u-M^{2}_{\tilde{e}})} \right)
\right]. \nonumber \\\label{eq:1}
\end{eqnarray}

Since the photino is not actually a mass eigenstate\textemdash
for our calculation is carried out in the context of
SQED\textemdash, we have assigned the neutralino mass to the
photino. In any case, the electron mass could be neglected compared
with the sparticle masses, and Eq.~(\ref{eq:1}) simplifies to
\begin{eqnarray}
\frac{d \sigma}{d \Omega}(e^{-}e^{+}\rightarrow \tilde{\gamma}\tilde{\gamma})&=&
\frac{\alpha^{2}}{4s} \sqrt{1- \left( \frac{2m_{\tilde{\gamma}}}{\sqrt{s}} \right)^{2}}
\left[
\left( \frac{t-m_{\tilde{\gamma}}^{2}}{t-M^{2}_{\tilde{e}}} \right)^{2}+
\left( \frac{u-m_{\tilde{\gamma}}}{u-M^{2}_{\tilde{e}}} \right)^{2}\right. \nonumber \\
&-& \left.
\frac{2sm^{2}_{\tilde{\gamma}}}{(u-M^{2}_{\tilde{e}})(t-M^{2}_{\tilde{e}})} \right],
\label{res1}
\end{eqnarray}
which is the result in \cite{kane,Dawson}. In addition, with
$m_{\tilde{\gamma}}=0$ we obtain Eq.~(\ref{fayetresult}), the result
derived by Fayet \cite{Fayet:1982ky}.

The total cross section for the process $e^{-}e^{+}\to \tilde{\gamma}\tilde{\gamma}$ is given by \cite{Dawson}
\begin{eqnarray}
\sigma (e^{-}e^{+}\to \tilde{\gamma}\tilde{\gamma})&=& \frac{2 \pi \alpha^{2}}{s^{2}} \left\{
 {\cal S}+2 \Delta \Lambda +
\frac{{\cal S}\Delta^{2}}{m^{4}_{\tilde{\gamma}}+M^{4}_{\tilde{e}}+M^{2}_{\tilde{e}}(s-2m^{2}_{\tilde{\gamma}})} +2
\frac{m^{2}_{\tilde{\gamma}}s \Lambda)}{s+2 \Delta} \right\}, \nonumber \\
\label{totcross}
\end{eqnarray}
where
\begin{eqnarray}
{\cal S}&=&\sqrt{s(s-4m^{2}_{\tilde{\gamma}})}, \nonumber \\
\Delta&=&M^{2}_{\tilde{e}}-m^{2}_{\tilde{\gamma}}, \nonumber \\
\Lambda&=&\ln \left[ \frac{s+2 \Delta- {\cal S}}{s+2 \Delta+ {\cal S}} \right].
\end{eqnarray}

The so-called International Linear Collider (ILC) will provide
opportunities for both discovery and precision
measurements~\cite{ilc}.  With the construction of the next generation
of $e^+e^-$ linear colliders, with a center-of-mass energy up to 1.5
TeV, capable of operating also in the $\gamma\gamma$, $\gamma
e^-$ and $e^-e^-$ modes, new perspectives arise for detecting new
physics beyond the standard model in processes having non-zero initial
electric charge (and non-zero lepton number).

If the photino is stable, then all supersymmetric cascade processes
ultimately decay into photinos. The photino is unseen as it leaves the
detector, and its existence can only be inferred by looking for
unbalanced momentum in a detector. In this way, it is
phenomenologically similar to neutrinos. The events produced by the
photinos display large discrepancy in energy and
momentum between the visible initial- and final-state
particles. Nowadays, this is the signature of
the LSP, which depending on the scenario can be the lightest neutralino
($\tilde{\chi}^{0}_{1}$), the gravitino
(%the gravitino (symbol $\tilde{G}$) is
the supersymmetric partner of the graviton), or the lightest sneutrinos
$\tilde{\nu}_{1}$, i.~e., the supersymmetric partners of neutrinos.  On
the other hand, in certain minimal Supergravity (mSUGRA) scenarios the LSP
can be a light photino (it means that $\tilde{\chi}^{0}_{1} \approx
\tilde{\gamma}$) \cite{sps1,sps2,sps} with an acceptable cosmological
abundance \cite{Olive:1994qq}. 

In the following we shall present our numerical results assuming
the photino to be the lightest neutralino of the MSSM. In the SPS
scenarios, its mass ranges between 70\, GeV and 200\, GeV, whereas the
selectron mass ranges from 200 GeV to 1.5 TeV (see
Table~\ref{tab:tmasses}).

\begin{figure*}[t]
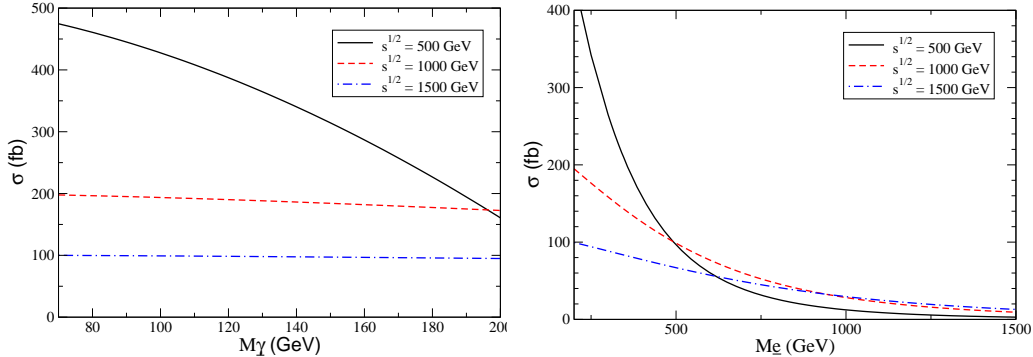

\begin{center}
%\begin{tabular}[cc]
\includegraphics[scale=0.28]{totfotinomfotino.eps} 
\includegraphics[scale=0.28]{totfotinomseletron.eps}
%\end{tabular}
\caption{Total cross section of photino production in $e^{-}e^{+}$
  collisions for different CM energies: $\sqrt{s}=0.5$ TeV (solid
  line), for $\sqrt{s}=1.0$ TeV (dashed line) and for $\sqrt{s}=1.5$
  TeV (dot-dashed line).  Left panel: Cross section as a function of
  the photino mass, for $M_{\tilde{e}}=400$ GeV. Right panel: Cross
  section as a function of the selectron mass, for
  $M_{\tilde{\gamma}}=79$ GeV.}
\label{fot2}
\end{center}
\end{figure*}

In Fig.~\ref{fot2} we show results for the total cross section of
photino production as a function of the photino and selectron masses,
for three different CM energies.  To study the dependence on the
photino mass, we fix %$M_{\tilde{e}}=400$ GeV
$M_{\tilde{e}}=202$ GeV for the selectron mass (SPS1a scenario).  The
cross sections for $\sqrt{s}=0.5$ TeV are largest, the results for
$\sqrt{s}=1$ TeV and $\sqrt{s}=1.5$ TeV being insensitive to the
photino mass.  For the dependency on the selectron mass, we
use %$M_{\tilde{\gamma}}=79$ GeV
$M_{\tilde{\gamma}}=96$ GeV for the photino mass. For
$M_{\tilde{e}}>800$ GeV the cross sections for $\sqrt{s}=1$ TeV and
$\sqrt{s}=1.5$ TeV are very close to each other, but the results
differ for lighter selectron masses. For $\sqrt{s}=0.5$ TeV the cross
section grows up faster than in the previous cases for decreasing
selectrom masses. In conclusion, if the selectron mass is not much
heavier than $M_{\tilde{e}}\simeq 500$ GeV, the International Linear
Collider (ILC) with $\sqrt{s}=0.5$ TeV is likely to discover the
lightest neutralinos and to place constraints on the selectron mass.

Let us now estimate the number of photinos that will be produced in
the future ILC. Given the expected luminosity of ${\cal{L}}=1.5\times
10^{34}$\,cm$^{-2}$s$^{-1}$\cite{Brau:2012zz} for the 0.5 TeV mode, in one
year ($10^{7}$ s) 65,218 photinos will be produced in the more
optimistic SPS1 scenario, and 525 photinos in the less optimistic SPS2
scenario.

\section{Gluino Production in the MSSM}
\label{sec:gluinos}

Gluino and squark production in hadron colliders dominantly occurs via
strong interactions. Thus, their production rate may be expected to be
considerably larger than for sparticles with electroweak
interactions only, whose production has been studied in the literature.

The cross sections for the production of gluinos and squarks in hadron
collisions were calculated at the Born level quite some time ago
\cite{Dawson} and in the next-to-leading order (NLO) accuracy more
recently \cite{Zerwas} .  In the present study, as another example of
detailed calculation, we consider the inclusive production of gluinos
in $pp$ collisions. For more detailed descriptions of the procedures
see \cite{Mariotto:2008zt,danusa}.

We do not consider in detail top the squark production, for which our
assumptions are invalid, which calls for more involved
treatment~\cite{plehn}. In the following, we detail the steps necessary
to calculate the various contributing subprocesses.

\subsection{Subprocess $\bar{q}q \to \tilde{g}\tilde{g}$.}
\label{qqbsgsg}

\begin{figure}[tb]
\begin{center}
\epsfig{file=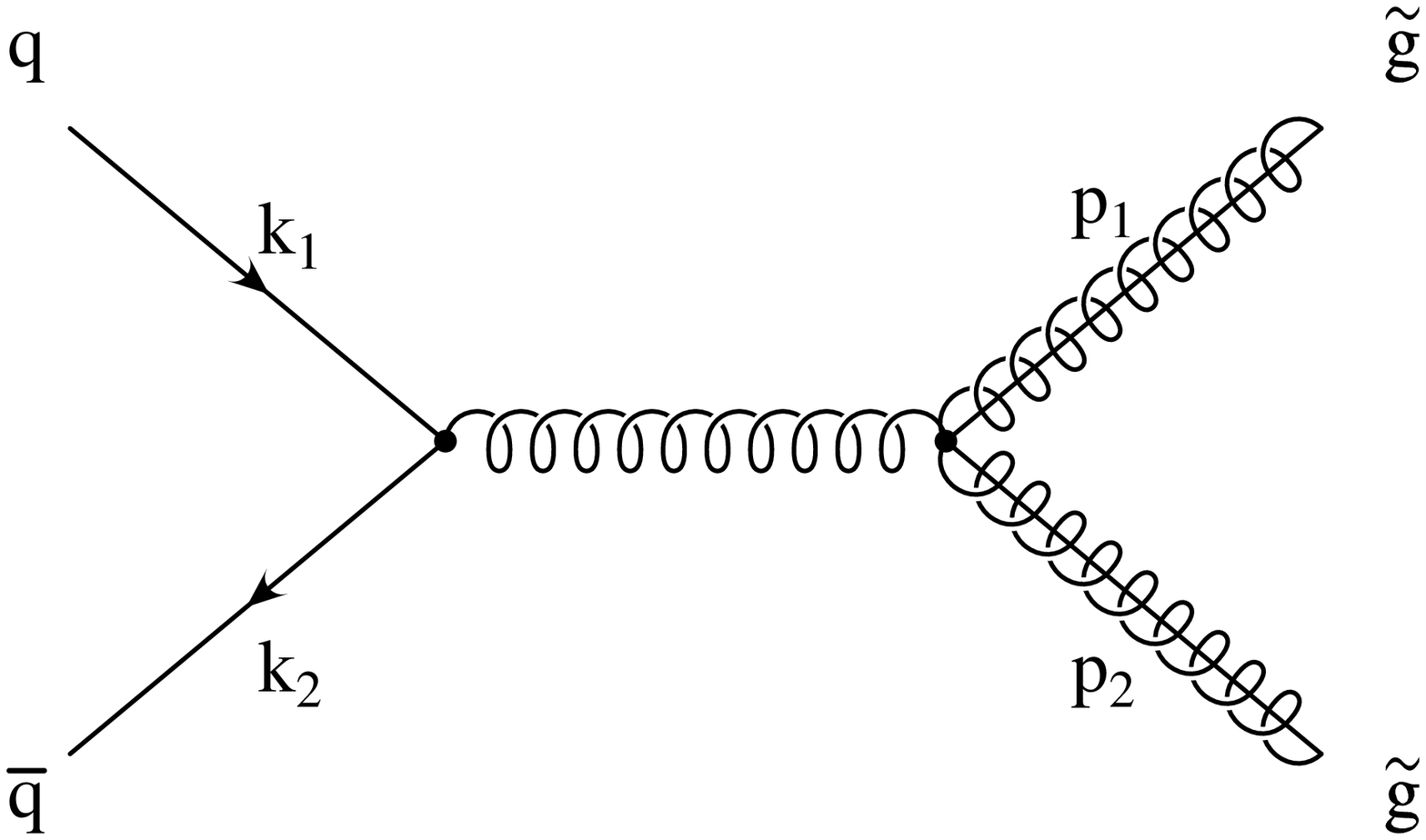,width=4cm}
\epsfig{file=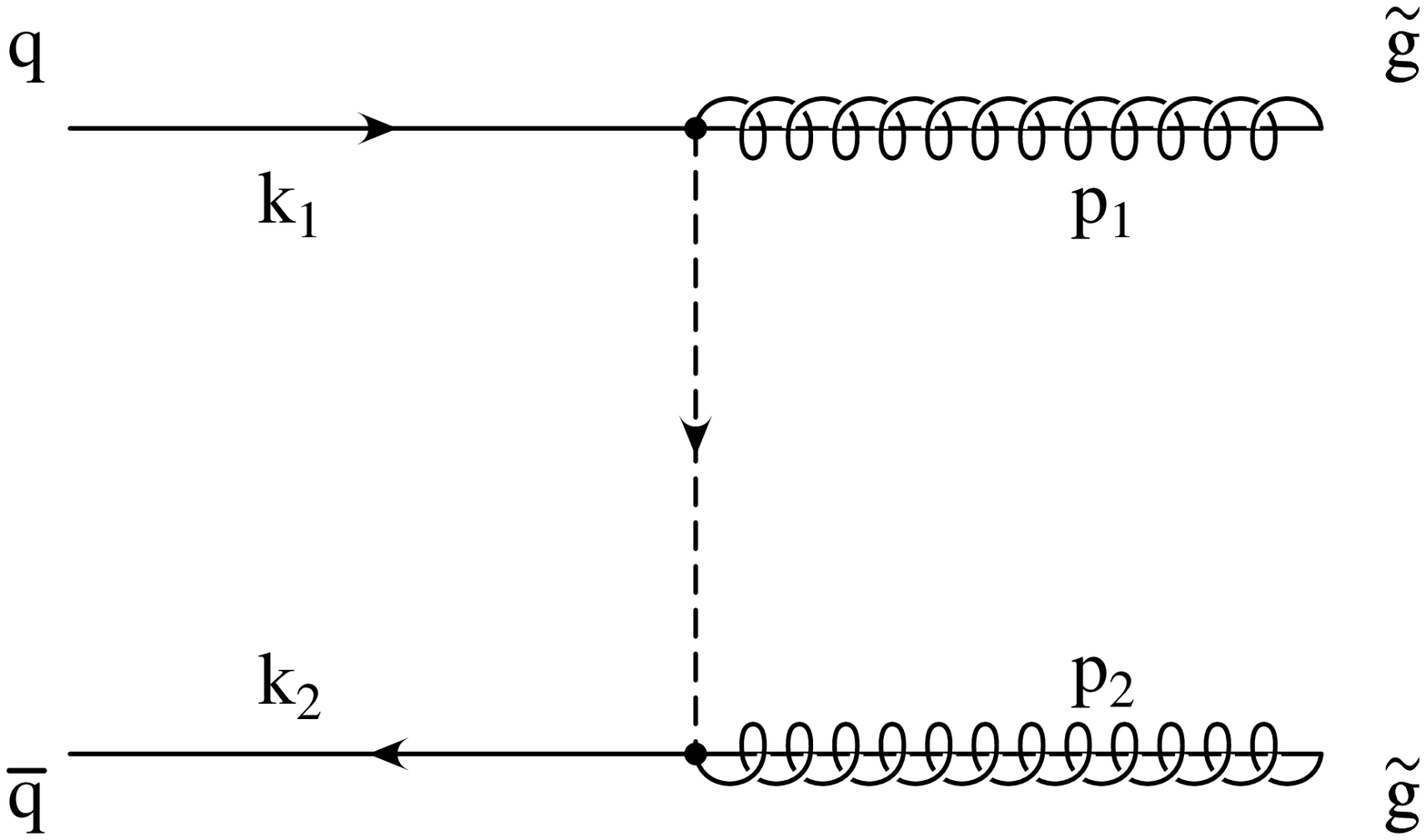,width=4cm}
\epsfig{file=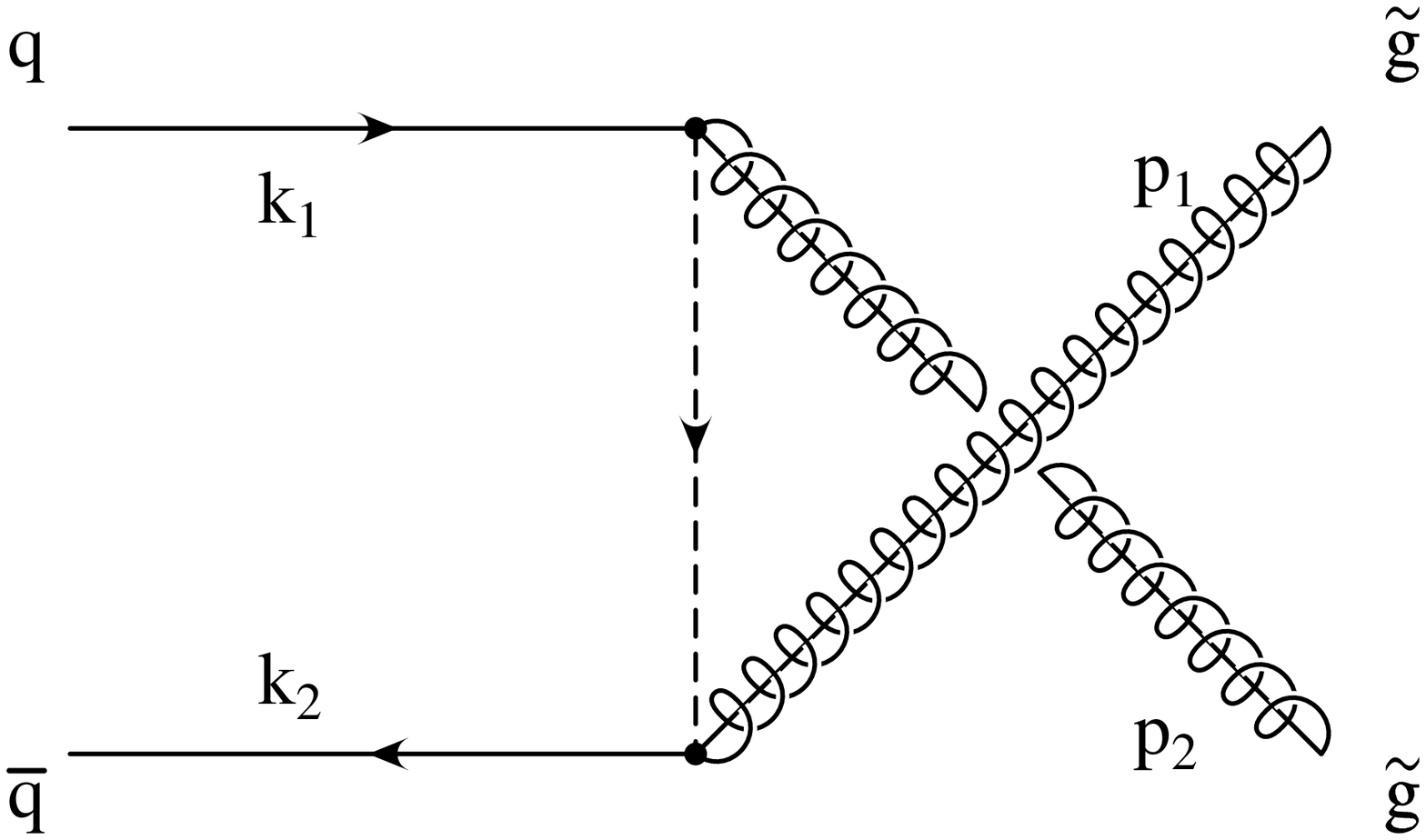,width=4cm}
\end{center}
\caption{Feynman diagrams for gluino pair production quark-antiquark initial states.}
\label{fig:QQB}
\end{figure}

The Feynman diagrams for gluino pair production coming from quark-antiquark initial 
states are drawn in Fig.~\ref{fig:QQB}. We denote the initial-state quark and
anti-quark momenta, spin and color by ($k_{1},s_{1},a$) and ($k_{2},s_{2},b$), and 
the final-state gluino momenta and spin by ($P_{1},s_{3},e$) and 
($P_{2},s_{4},d$), respectively.  

The Mandelstam variables are
\begin{eqnarray}
s&=&(k_{1}+k_{2})^{2}=(P_{1}+P_{2})^{2}=
2k_{1}\cdot k_{2}=2m^{2}_{\tilde{g}}+2P_{1}\cdot P_{2}=4E^{2}, 
\nonumber \\
t&=&(k_{1}-P_{1})^{2}=(P_{2}-k_{2})^{2}=m^{2}_{\tilde{g}}-2k_{1}\cdot P_{1}=
m^{2}_{\tilde{g}}-2k_{2}\cdot P_{2} \nonumber \\ &=&
m^{2}_{\tilde{g}}-2E^{2}+2E\sqrt{E^{2}-m^{2}_{\tilde{g}}}\cos \theta , \nonumber \\
u&=&(k_{1}-P_{2})^{2}=(P_{1}-k_{2})^{2}=m^{2}_{\tilde{g}}-2k_{1}\cdot P_{2}=
m^{2}_{\tilde{g}}-2k_{2}\cdot P_{1} \nonumber \\ &=&
m^{2}_{\tilde{g}}-2E^{2}-2E\sqrt{E^{2}-m^{2}_{\tilde{g}}}\cos \theta , \nonumber \\
s&+&t+u=2m^{2}_{\tilde{g}},
\end{eqnarray}
where $E$ is the center-of-mass energy of the colliding quarks, 
$m_{\tilde{g}}$ is the gluino mass and $\theta$ is the scattering angle.

The expressions for the amplitudes in these subprocesses are
\begin{eqnarray}
{\cal M}_{s}&=&(- \imath g_{s})
\left( w^{\dagger}(b)\bar{v}(k_{2},s_{2})T^{c}\gamma_{m}w(a)u(k_{1},s_{1}) \right)
\left( \frac{g^{mn}\delta^{cg}}{s} \right)(- \imath g_{s}) \cdot \nonumber \\
&\cdot&
\left( \Omega_{g}^{\dagger}(d)
\bar{u}(P_{2},s_{4})f^{deg}\gamma_{n}\Omega_{g}(e)u(P_{1},s_{3}) \right) 
\nonumber \\
&=&- \frac{g^{2}_{s}}{s} \left( \bar{v}(k_{2},s_{2}) \gamma_{m} u(k_{1},s_{1}) \right) 
\left( \bar{u}(P_{2},s_{4}) \gamma^{m}u(P_{1},s_{3}) \right) \left( w^{\dagger}(b)T^{c}w(a) \right) \nonumber \\
& \cdot & 
\left( f^{dec} \Omega_{g}^{\dagger}(d) \Omega_{g}(e) \right) , \nonumber \\
{\cal M}_{t}&=&(- \imath \sqrt{2} g_{s}(L-R))
\left( w^{\dagger}(b)\bar{v}(k_{2},s_{2})T^{c} \Omega_{c}(d)u(P_{2},s_{4}) \right) \nonumber \\
& \cdot &
\left( \frac{\imath \delta^{cf}}{t-M^{2}_{\tilde{q}}} \right)
(- \imath \sqrt{2} g_{s}(L-R)) 
\left( \Omega_{f}^{\dagger}(e)\bar{u}(P_{1},s_{3})T^{f}w(a)u(k_{1},s_{1}) \right) \nonumber \\
&=&- \frac{2 \imath g^{2}_{s}}{t-M^{2}_{\tilde{q}}} 
\left( \bar{v}(k_{2},s_{2})(L-R) u(P_{2},s_{4}) \right) 
\left( \bar{u}(P_{1},s_{4})(L-R) u(k_{1},s_{1}) \right) \nonumber \\
& \cdot & 
\left( w^{\dagger}(b)T^{c}\Omega_{c}(d) \right) 
\left( \Omega_{c}^{\dagger}(e)T^{c}w(a) \right)   \nonumber \\
&=&- \frac{2 \imath g^{2}_{s}}{t-M^{2}_{\tilde{q}}}(L+R)^{2} 
\left( \bar{v}(k_{2},s_{2})u(P_{2},s_{4}) \right) 
\left( \bar{u}(P_{1},s_{4})u(k_{1},s_{1}) \right) \nonumber \\
& \cdot & 
\left( w^{\dagger}(b)T^{c}\Omega_{c}(d) \right) 
\left( \Omega_{c}^{\dagger}(e)T^{c}w(a) \right)   \nonumber \\
&=&- \frac{4 \imath g^{2}_{s}}{t-M^{2}_{\tilde{q}}}
\left( \bar{v}(k_{2},s_{2})u(P_{2},s_{4}) \right) 
\left( \bar{u}(P_{1},s_{4})u(k_{1},s_{1}) \right)  
\left( w^{\dagger}(b)T^{c}\Omega_{c}(d) \right) 
\left( \Omega_{c}^{\dagger}(e)T^{c}w(a) \right) ,   \nonumber \\
{\cal M}_{u}&=&-(- \imath \sqrt{2} g_{s}(L-R)) (- \imath \sqrt{2} g_{s}(L-R)) 
\left( w^{\dagger}(b)\bar{v}(k_{2},s_{2})T^{c} \Omega_{c}(e)u(P_{1},s_{3}) \right)
\nonumber \\
& \cdot &
\left( \frac{\imath \delta^{cf}}{u-M^{2}_{\tilde{q}}} \right)
\left( \Omega_{f}^{\dagger}(d)\bar{u}(P_{2},s_{4})T^{f}w(a)u(k_{1},s_{1}) \right) 
 \nonumber \\
&=&+ \frac{4 \imath g^{2}_{s}}{t-M^{2}_{\tilde{q}}}
\left( \bar{v}(k_{2},s_{2})u(P_{1},s_{3}) \right) 
\left( \bar{u}(P_{2},s_{4})u(k_{1},s_{1}) \right)  
\left( w^{\dagger}(b)T^{c}\Omega_{c}(e) \right) 
\left( \Omega_{c}^{\dagger}(d)T^{c}w(a) \right) , \nonumber \\
\label{ampqqbsgsgcu}
\end{eqnarray}
where $w(a)$ and $\Omega(a)$ are the color wavefunctions of the quarks
and gluinos, respectively \cite{ait1}. 

The $\Omega(a)$'s are $8 \times 8$ matrices, in the octet
representations of $SU(3)_{C}$ group, satisfying the following
relations
\begin{eqnarray}
\left( \Omega(a) \right)_{bc}&=&- \imath f_{abc}, \nonumber \\
\left[ \Omega(a), \Omega(b) \right]&=&\imath f_{abc}\Omega(c).  
\end{eqnarray}

The total amplitude is given by
\begin{equation}
{\cal M}={\cal M}_{s}+{\cal M}_{t}+{\cal M}_{u}.
\label{faby1}
\end{equation}
%In order to do more easily the calculation we will define the convenient colour factors:
Defining the appropriate color factors
\begin{eqnarray}
G_{s} &\equiv& \frac{g_{s}^{2}}{s}\left( w^{\dagger}(b)T^{g}w(a) \right) 
\left(f^{deg} \Omega_{g}^{\dagger}(d) \Omega_{g}(e) \right) , \nonumber \\
G_{t} &\equiv& \frac{g_{s}^{2}}{t-M^{2}_{\tilde{q}}}
\left( w^{\dagger}(b)T^{c}\Omega_{c}(d) \right) 
\left( \Omega_{c}^{\dagger}(e)T^{c}w(a) \right) , \nonumber \\
G_{u} &\equiv& \frac{g_{s}^{2}}{u-M^{2}_{\tilde{q}}}
\left( w^{\dagger}(b)T^{c}\Omega_{c}(e) \right) 
\left( \Omega_{c}^{\dagger}(d)T^{c}w(a) \right)  ,
\label{faby4}
\end{eqnarray}
and the following flavor factors
\begin{eqnarray}
S_{s}&=&\left( \bar{v}(k_{2},s_{2}) \gamma_{m} u(k_{1},s_{1}) \right) 
\left( \bar{u}(P_{2},s_{4}) \gamma^{m}u(P_{1},s_{3}) \right) , \nonumber \\
S_{t}&=&\left( \bar{v}(k_{2},s_{2})u(P_{2},s_{4}) \right) 
\left( \bar{u}(P_{1},s_{3})u(k_{1},s_{1}) \right) , \nonumber \\
S_{u}&=&\left( \bar{v}(k_{2},s_{2})u(P_{1},s_{3}) \right) 
\left( \bar{u}(P_{2},s_{4})u(k_{1},s_{1}) \right) , \nonumber
\label{faby3}
\end{eqnarray}
we can rewrite the amplitude (\ref{faby1}) in the form
\begin{equation}
{\cal M}=S_{s}G_{s}+ \imath S_{t}G_{t}- \imath S_{u}G_{u}\,.
\label{faby2}
\end{equation}

Squaring this amplitude and summing over initial and final color and
spins, we are led to the expression
\begin{eqnarray}
\sum_{s_{1}, s_{2}, s_{3}, s_{4}}|{\cal M}|^{2}&=& |S_{s}|^{2}|G_{s}|^{2}+ \imath
S^{*}_{s}S_{t}G^{*}_{s}G_{t}- \imath S^{*}_{s}S_{u}G^{*}_{s}G_{u}- \imath S_{s}S^{*}_{t}G_{s}G^{*}_{t}+ 
|S_{t}|^{2}|G_{t}|^{2} \nonumber \\
&-&S^{*}_{t}S_{u}G^{*}_{t}G_{u}+ \imath S_{s}S^{*}_{u}G_{s}G^{*}_{u}
+|S_{u}|^{2}|G_{u}|^{2}\,.
\label{faby3}
\end{eqnarray}

Let us first examine the flavor factors. The usual quantum
field-theory techniques yield the equalities
\begin{eqnarray}
\sum_{s_{1}, s_{2}, s_{3}, s_{4}}|S_{s}|^{2}&=& \mbox{Tr} \left[ 
\not\hspace{-.4ex}{k}_{2}\gamma_{m}\not\hspace{-.4ex}{k}_{1}\gamma_{n} \right] \,\ 
\mbox{Tr}\left[ ( \not\hspace{-.4ex}{P}_{2}+m_{\tilde{g}})\gamma^{m}
( \not\hspace{-.4ex}{P}_{1}+m_{\tilde{g}})\gamma^{n} \right] \nonumber \\
&=&k^{o}_{2}k^{p}_{2}\mbox{Tr} 
\left[ \gamma_{o}\gamma_{m}\gamma_{p}\gamma_{n} \right]%\nonumber \\ &+&
+ \mbox{Tr}\left[ 
\not\hspace{-.4ex}{P}_{2}\gamma^{m}\not\hspace{-.4ex}{P}_{1}\gamma^{n}+
m^{2}_{\tilde{g}}\gamma^{m}\gamma^{n}\right. \nonumber \\ 
&+& \left.  m_{\tilde{g}}\left( 
\not\hspace{-.4ex}{P}_{2}\gamma^{m}\gamma^{n}+ 
\gamma^{m}\not\hspace{-.4ex}{P}_{1}\gamma^{n} \right) \right] \nonumber \\
&=&\left[ 4 \cdot \left(k_{2m}k_{1n}-g_{mn} k_{1} \cdot k_{2}+k_{2n}k_{1m}
\right)\right] \cdot
\left[ 4 \cdot \left(P_{2}^{m}P_{1}^{n}-g^{mn} P_{1} \cdot P_{2} \right. \right. 
\nonumber \\ &+& \left. \left.
P_{2}^{n}P_{1}^{m}+m^{2}_{\tilde{g}}g^{mn}
\right)\right]  \nonumber \\
&=&32 \left[ \left( k_{1} \cdot  P_{1} \right) \left( k_{2} \cdot  P_{2} \right) 
+ \left( k_{1} \cdot  P_{2} \right) \left( k_{2} \cdot  P_{1} \right) +
m^{2}_{\tilde{g}} \left( k_{1} \cdot  k_{2}+2m^{2}_{\tilde{g}} \right) 
\right] \nonumber \\
&=&8 \left[ \left( m^{2}_{\tilde{g}}-t \right)^{2}+
\left( m^{2}_{\tilde{g}}-u \right)^{2}+2sm^{2}_{\tilde{g}} \right] , \nonumber \\
\sum_{s_{1}, s_{2}, s_{3}, s_{4}}S^{*}_{s}S_{t}&=&S^{*}_{t}S_{s}= \mbox{Tr} \left[ 
\not\hspace{-.4ex}{k}_{2}\gamma_{m}\not\hspace{-.4ex}{k}_{1}
( \not\hspace{-.4ex}{P}_{1}+m_{\tilde{g}})\gamma^{m}
( \not\hspace{-.4ex}{P}_{2}+m_{\tilde{g}}) \right] \nonumber \\
&=& \mbox{Tr} \left[ 
\not\hspace{-.4ex}{k}_{2}\gamma_{m}\not\hspace{-.4ex}{k}_{1}
\not\hspace{-.4ex}{P}_{1}\gamma^{m}\not\hspace{-.4ex}{P}_{2}+
m_{\tilde{g}}\not\hspace{-.4ex}{k}_{2}\gamma_{m}\not\hspace{-.4ex}{k}_{1}
\not\hspace{-.4ex}{P}_{1}\gamma^{m}+
m_{\tilde{g}}\not\hspace{-.4ex}{k}_{2}\gamma_{m}\not\hspace{-.4ex}{k}_{1}\gamma^{m}
\not\hspace{-.4ex}{P}_{2} \right. \nonumber \\ 
&+& \left. m^{2}_{\tilde{g}}
\not\hspace{-.4ex}{k}_{2}\gamma_{m}\not\hspace{-.4ex}{k}_{1}\gamma^{m}
\right] \nonumber \\
&=&16 \left( k_{1}\cdot P_{1} \right) \left( k_{2}\cdot P_{2} \right) +
m^{2}_{\tilde{g}} \left( k_{1}\cdot k_{2} \right) \nonumber \\
&=&4 \left( m^{2}_{\tilde{g}}-t \right) +m^{2}_{\tilde{g}}s, \nonumber \\ 
\sum_{s_{1}, s_{2}, s_{3}, s_{4}}S^{*}_{s}S_{u}&=&S^{*}_{u}S_{s}= \mbox{Tr} \left[ 
\not\hspace{-.4ex}{k}_{2}\gamma_{m}\not\hspace{-.4ex}{k}_{1}
( \not\hspace{-.4ex}{P}_{2}+m_{\tilde{g}})\gamma^{m}
( \not\hspace{-.4ex}{P}_{1}+m_{\tilde{g}}) \right] \nonumber \\
&=& \mbox{Tr} \left[ 
\not\hspace{-.4ex}{k}_{2}\gamma_{m}\not\hspace{-.4ex}{k}_{1}
\not\hspace{-.4ex}{P}_{2}\gamma^{m}\not\hspace{-.4ex}{P}_{1}+
m_{\tilde{g}}\not\hspace{-.4ex}{k}_{2}\gamma_{m}\not\hspace{-.4ex}{k}_{1}
\not\hspace{-.4ex}{P}_{2}\gamma^{m}+
m_{\tilde{g}}\not\hspace{-.4ex}{k}_{2}\gamma_{m}\not\hspace{-.4ex}{k}_{1}\gamma^{m}
\not\hspace{-.4ex}{P}_{1}\right. \nonumber \\ 
&+& \left. m^{2}_{\tilde{g}}
\not\hspace{-.4ex}{k}_{2}\gamma_{m}\not\hspace{-.4ex}{k}_{1}\gamma^{m}
\right] \nonumber \\
&=&16 \left( k_{1}\cdot P_{2} \right) \left( k_{2}\cdot P_{1} \right) +
m^{2}_{\tilde{g}} \left( k_{1}\cdot k_{2} \right) \nonumber \\
&=&4 \left( m^{2}_{\tilde{g}}-u \right) +m^{2}_{\tilde{g}}s, \nonumber \\
\sum_{s_{1}, s_{2}, s_{3}, s_{4}}|S_{t}|^{2}&=& \mbox{Tr} \left[ 
\not\hspace{-.4ex}{k}_{2}( \not\hspace{-.4ex}{P}_{2}+m_{\tilde{g}}) \right] 
\,\ \mbox{Tr}\left[ 
( \not\hspace{-.4ex}{P}_{1}+m_{\tilde{g}}) \not\hspace{-.4ex}{k}_{1} 
\right] \nonumber \\
&=& \mbox{Tr} \left[ \not\hspace{-.4ex}{k}_{1} \not\hspace{-.4ex}{P}_{1} \right]
+ \mbox{Tr} \left[ \not\hspace{-.4ex}{k}_{2} \not\hspace{-.4ex}{P}_{2} \right] 
+m_{\tilde{g}} \left\{ \mbox{Tr} \left[ \not\hspace{-.4ex}{k}_{1} \right] +
\mbox{Tr} \left[ \not\hspace{-.4ex}{k}_{2} \right] \right\}  \nonumber \\
&=&4 \left[ P_{1} \cdot k_{1}+P_{2} \cdot k_{2} \right] \nonumber \\
&=&4 \left( m^{2}_{\tilde{g}}-t \right) , \nonumber \\
\sum_{s_{1}, s_{2}, s_{3}, s_{4}}S^{*}_{s}S_{u}&=&S^{*}_{u}S_{s}= 
\mbox{Tr} \left[ \not\hspace{-.4ex}{k}_{1} \not\hspace{-.4ex}{P}_{1} \right]
\mbox{Tr} \left[ \not\hspace{-.4ex}{k}_{2} \not\hspace{-.4ex}{P}_{1} \right] -
\mbox{Tr} \left[ \not\hspace{-.4ex}{k}_{1} \not\hspace{-.4ex}{P}_{2} \right]
\mbox{Tr} \left[ \not\hspace{-.4ex}{k}_{2} \not\hspace{-.4ex}{P}_{2} \right] +
m^{2}_{\tilde{g}}
\mbox{Tr} \left[ \not\hspace{-.4ex}{k}_{1} \not\hspace{-.4ex}{k}_{2} \right] 
 \nonumber \\
&=&4 \left[ \left( k_{1}\cdot P_{1} \right) \left( k_{2}\cdot P_{1} \right) -
\left( k_{1}\cdot P_{2} \right) \left( k_{2}\cdot P_{2} \right) +
m^{2}_{\tilde{g}} \left( k_{1}\cdot k_{2} \right) \right]  \nonumber \\
&=&8sm^{2}_{\tilde{g}}, \nonumber \\
\sum_{s_{1}, s_{2}, s_{3}, s_{4}}|S_{u}|^{2}&=& \mbox{Tr} \left[ 
( \not\hspace{-.4ex}{P}_{2}+m_{\tilde{g}}) \not\hspace{-.4ex}{k}_{1} \right] 
\,\ \mbox{Tr}\left[ 
( \not\hspace{-.4ex}{P}_{1}+m_{\tilde{g}}) \not\hspace{-.4ex}{k}_{2} \right] 
 \nonumber \\
&=& \mbox{Tr} \left[ \not\hspace{-.4ex}{k}_{1} \not\hspace{-.4ex}{P}_{2} \right]
+ \mbox{Tr} \left[ \not\hspace{-.4ex}{k}_{2} \not\hspace{-.4ex}{P}_{1} \right] 
+m_{\tilde{g}} \left\{ \mbox{Tr} \left[ \not\hspace{-.4ex}{k}_{1} \right] +
\mbox{Tr} \left[ \not\hspace{-.4ex}{k}_{2} \right] \right\}  \nonumber \\
&=&4 \left[ P_{2} \cdot k_{1}+P_{1} \cdot k_{2} \right]  \nonumber \\
&=&4 \left( m^{2}_{\tilde{g}}-u \right) .
\label{faby5}
\end{eqnarray}

Consider next the color factors. Apart from the averaging factor (1/9), 
we get the following results:
\begin{eqnarray}
\sum_{a,b,d,e}|G_{s}|^{2} &=&\frac{g_{s}^{4}}{4s^{2}} \mbox{Tr}\left[ \lambda^{c}\lambda^{g} \right]
f^{dec}f^{dec^{\prime}}f^{deg}f^{deg^{\prime}}
\nonumber \\
&=&\frac{g_{s}^{4}}{4s^{2}}\cdot 16 \cdot 72, \nonumber \\
\sum_{a,b,d,e}G^{*}_{s}G_{t} &=& \sum_{a,b,d,e}G_{s}G^{*}_{t}= 
\frac{g_{s}^{4}}{s \left( t-M^{2}_{\tilde{q}} 
\right)} \cdot 16 \cdot 72 , \nonumber \\
\sum_{a,b,d,e}G^{*}_{s}G_{u} &=& \sum_{a,b,d,e}G_{s}G^{*}_{u}= 
\frac{g_{s}^{4}}{s \left( u-M^{2}_{\tilde{q}} \right)} \cdot 16 \cdot 72 
, \nonumber \\
\sum_{a,b,d,e}|G_{t}|^{2} &=& 
\frac{g_{s}^{4}}{\left( t-M^{2}_{\tilde{q}} \right)^{2}}\cdot 16 \cdot 72 
\cdot \frac{3}{9}, \nonumber \\
\sum_{a,b,d,e}G^{*}_{t}G_{u} &=& \sum_{a,b,d,e}G_{t}G^{*}_{u}=
\frac{g_{s}^{4}}{\left( t-M^{2}_{\tilde{q}} \right) \left( u-M^{2}_{\tilde{q}} \right)} \cdot 16 \cdot 72 , \nonumber \\
\sum_{a,b,d,e}|G_{u}|^{2} &=&
\frac{g_{s}^{4}}{4 \left( u-M^{2}_{\tilde{q}} \right)^{2}}\cdot 16 \cdot 72 
\cdot \frac{3}{9} . 
%\nonumber \\
\label{faby6} 
\end{eqnarray}

We may now substitute Eqs.~(\ref{faby5},\ref{faby6}, \ref{faby3}) into
the usual expression for the  differential cross section,
\begin{eqnarray}
\frac{d\sigma}{dt} &=&
\frac{1}{16\pi s^2}
\left(
\frac{1}{64}\sum_{a,b,d,e}
\frac{1}{4}\sum_{s_{1}, s_{2}, s_{3}, s_{4}} |{\cal M}|^2\,
\right).
\end{eqnarray}

We have checked that our analytical calculations agree with the
results of the FeynArts program \cite{Hahn:2000kx}, with the MSSM code
\cite{Hahn:2001rv}. Finally, we find the following expression:
\begin{eqnarray}
  \frac{d \sigma}{d \hat{t}}( \bar{q}q \to \tilde{g}\tilde{g})&=& \hspace{- 1mm} \frac{8 \pi \alpha^{2}_{s}}{9 \hat{s}^{2}} 
  \left \{ \frac{4}{3} \left( 
      \frac{m^{2}_{\tilde{g}}- \hat{t}}{M^{2}_{\tilde{q}}- \hat{t}} \right)^{2} +
    \frac{4}{3} \left( \frac{m^{2}_{\tilde{g}}- \hat{u}}{M^{2}_{\tilde{q}}- \hat{u}} \right)^{2} \right. \nonumber \\
  &+& \left.
    \frac{3}{\hat{s}^{2}} \left[ (m^{2}_{\tilde{g}}- \hat{t})^{2}+(m^{2}_{\tilde{g}}- \hat{u})^{2}+
      2m^{2}_{\tilde{g}}\hat{s} \right] \right. \nonumber \\
  &-& \left. 3
    \left[ \frac{(m^{2}_{\tilde{g}}- \hat{t})^{2}+m^{2}_{\tilde{g}}\hat{s}}{\hat{s}(M^{2}_{\tilde{q}}- \hat{t})} \right]
    -3 
    \left[ \frac{(m^{2}_{\tilde{g}}- \hat{u})^{2}+m^{2}_{\tilde{g}}\hat{s}}{\hat{s}(M^{2}_{\tilde{q}}- \hat{u})} \right] \right. \nonumber \\
  &+& \left.
    \frac{1}{3} \frac{m^{2}_{\tilde{g}}\hat{s}}{(M^{2}_{\tilde{q}}- \hat{t})(M^{2}_{\tilde{q}}- \hat{u})} \right \}.
\label{formula1}
\end{eqnarray}
These results agree with those in Refs.~\onlinecite{tata,Dawson}.

\subsection{Subprocess $gg \to \tilde{g}\tilde{g}$.}
\label{ggsgsg}

\begin{figure}[tb]
\begin{center}
\epsfig{file=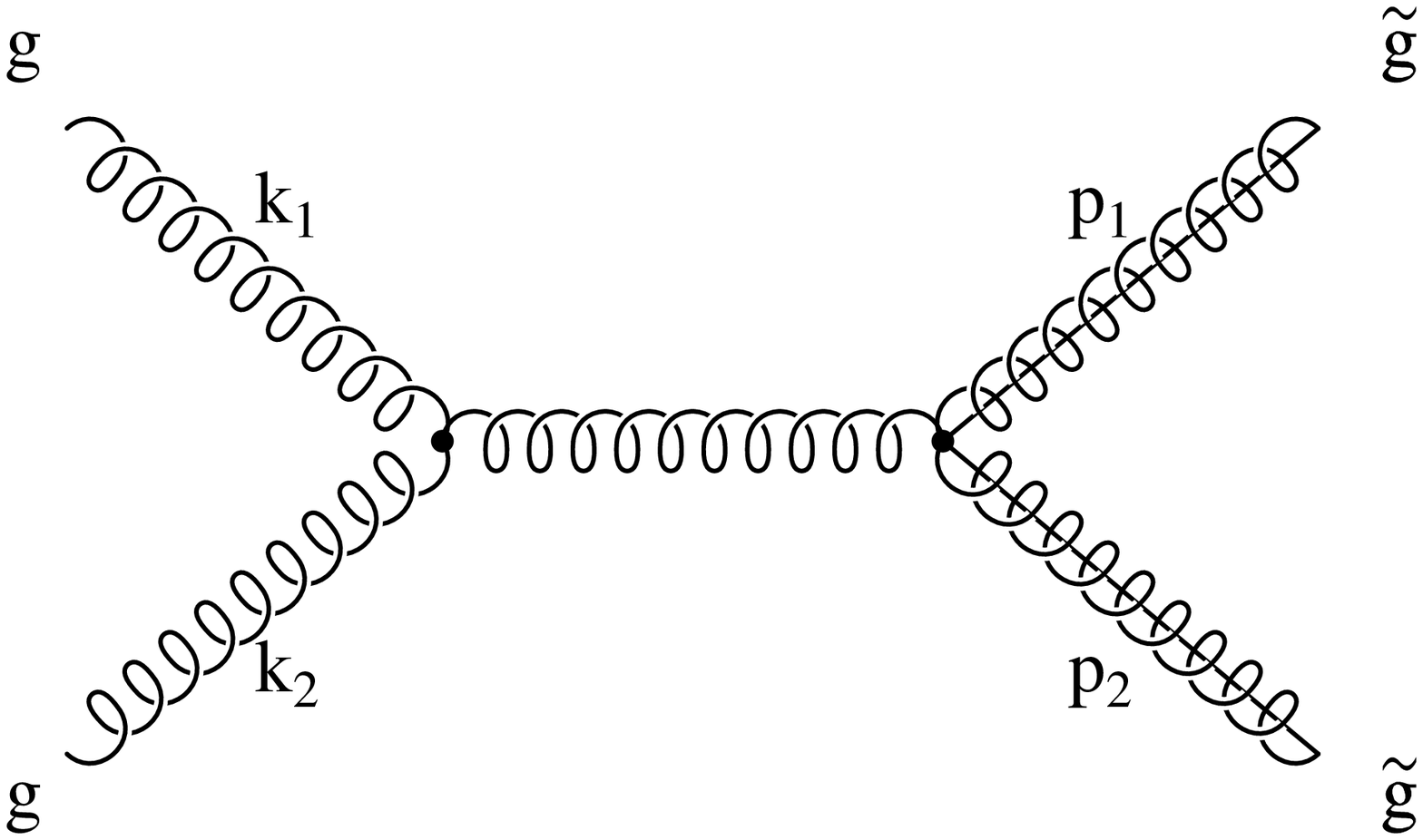,width=4cm}
\epsfig{file=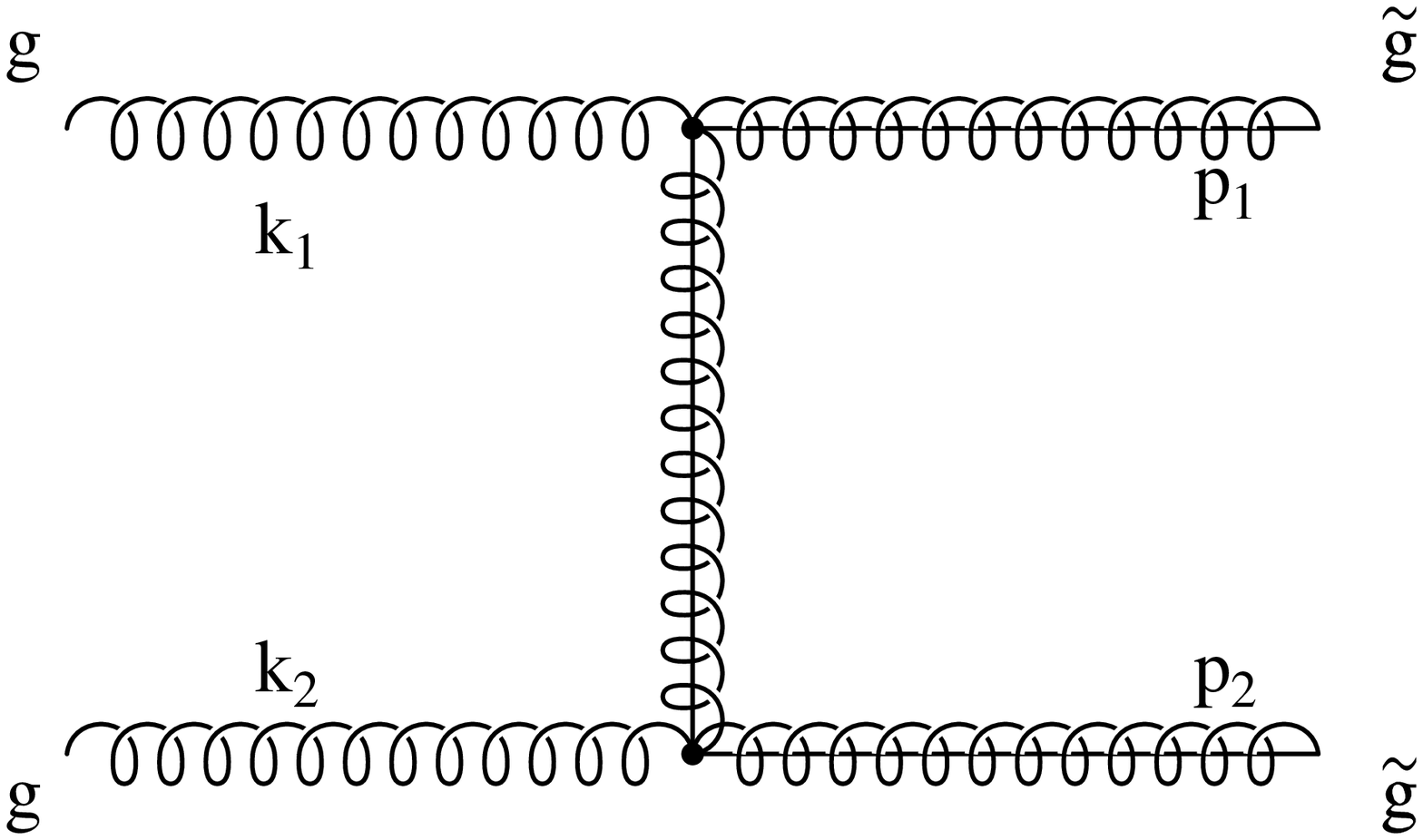,width=4cm}
\epsfig{file=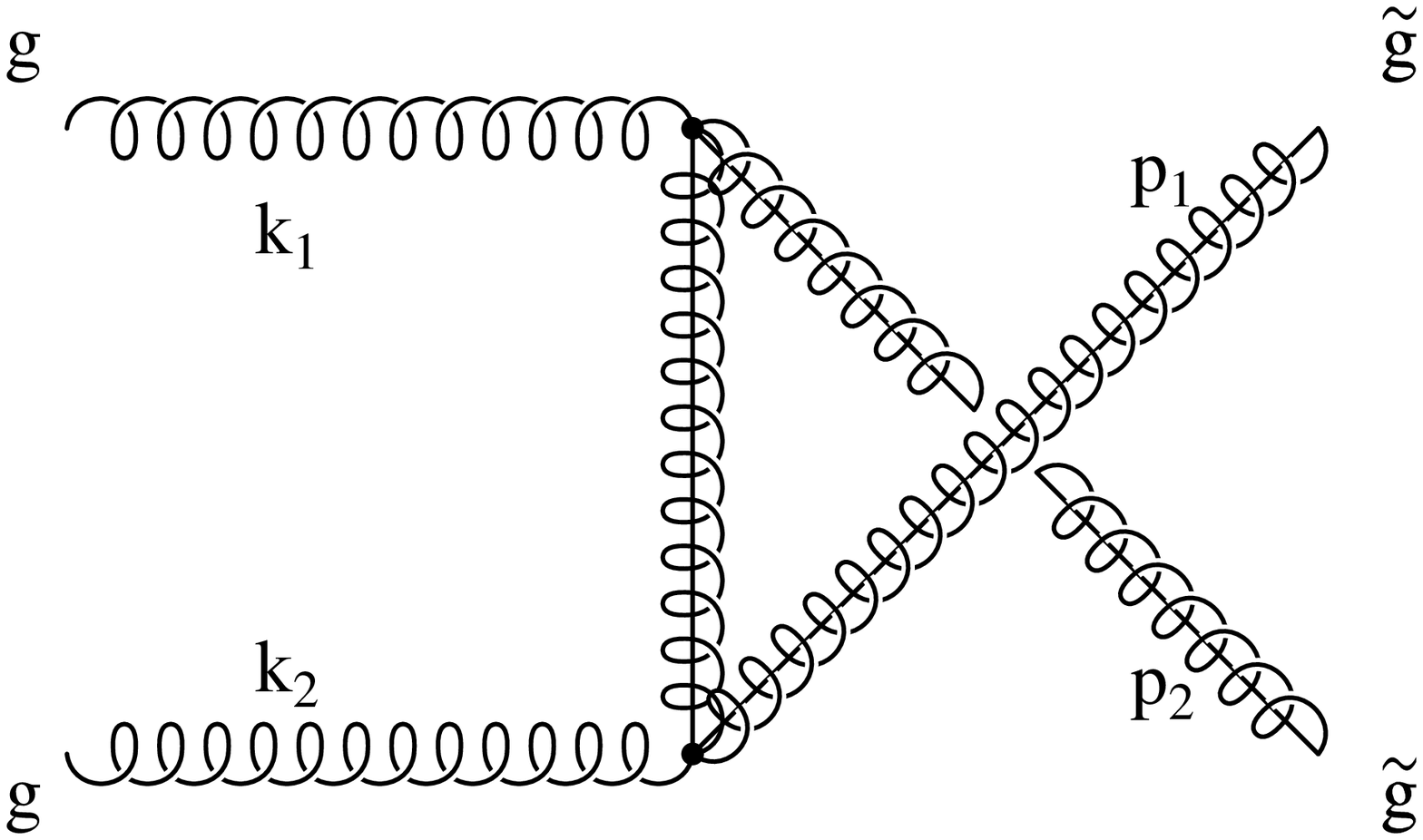,width=4cm}
\end{center}
\caption{Feynman diagrams for gluino pair production gluon-gluon initial states.}
\label{fig:GG}
\end{figure}

The Feynman diagrams for gluino production coming from gluon fusion
are depicted in Fig.~\ref{fig:GG}. The initial-state gluons have
$SU(3)_c$ adjoint representation indices $a$ and $b$, with momenta
$k_1$ and $k_2$ and polarization vectors
$\varepsilon_{1}^{m}(k_{1},\lambda_{1})$ and
$\varepsilon_{2}^{m}(k_{2},\lambda_2)$, respectively.  The final-state
gluinos carry adjoint representation indices $c$ and $d$, with momenta
$P_{3}$ and $P_{4}$.

The Feynman amplitudes are given by
\begin{eqnarray}
{\cal M}_{s}&=&(- \imath g_{s}f^{abe}) \left( \epsilon_{m}(k_{1})a^{a}(c) \epsilon^{*}_{n}(k_{2}) a^{b}(c) \right)
\left( \frac{g^{mn}\delta^{ef}}{s} \right) (- \imath g_{s}f^{cdf}) \left( \Omega^{\dagger}(c) \bar{u}(P_{1}) \Omega(d)u(P_{2}) \right) \nonumber \\
&\cdot&
\left[ g_{mn}(P_{1}-P_{2})_{r}+ g_{nr}(P_{2}+Q)_{m}- g_{rm}(Q+P_{1})_{r} \right]\,, \nonumber \\
{\cal M}_{t}&=&(- \imath g_{s}f^{bde}) \left( \epsilon_{m}(k_{2})a^{b}(c) u(P_{2}) \Omega(d) \right) 
\left( \frac{\imath 
( \not\hspace{-.1ex}{q}+m_{\tilde{g}})}{t-m^{2}_{\tilde{g}}}\delta^{de}g^{mn} \right) 
(- \imath g_{s}f^{bcf}) \nonumber \\
&\cdot& \left( \bar{u}(k_{1}) \Omega^{\dagger}(c) \epsilon_{n}(k_{1})a^{a}(c)  \right)\,, \nonumber \\
{\cal M}_{u}&=&(- \imath g_{s}f^{bce}) \left( \epsilon_{m}(k_{2})a^{b}(c) u(k_{1}) \Omega(c) \right) 
\left( \frac{\imath 
( \not\hspace{-.1ex}{q}+m_{\tilde{g}})}{u-m^{2}_{\tilde{g}}}\delta^{de}g^{mn} \right)
(- \imath g_{s}f^{bdf}) \nonumber \\
&\cdot& 
\left( \bar{u}(k_{2}) \Omega^{\dagger}(d)\epsilon_{n}(k_{1})a^{a}(c) \right).
\label{ampggsgsgcu}
\end{eqnarray}

Before presenting our results for this case, we find it interesting to
present a brief review of the polarization vectors used to describe
real photons as well as real gluons.  We choose to work with real
transverse polarization vectors $\varepsilon_1$ and $\varepsilon_2$,
both of which must be orthogonal to the initial-state collision
axis in the center-of-momentum frame.  We can hence write the
following relations:
\begin{eqnarray}
\varepsilon_{i} \cdot \varepsilon_{j} &=& - \delta_{ij}\,, \nonumber \\
\varepsilon_{1} \cdot p_{1} &=&
\varepsilon_{2} \cdot p_{1} =
\varepsilon_{1} \cdot p_{2} =
\varepsilon_{2} \cdot p_{2} = 0\,\,\,\, \mbox{(Lorentz condition)}, \nonumber \\
\varepsilon_{1} \cdot k_{2} &=& -\varepsilon_{1} \cdot k_{1}\,, \nonumber \\
%\qquad\qquad\quad
\varepsilon_{2} \cdot k_{2} &=& -\varepsilon_{2} \cdot k_{1} \,,
\end{eqnarray}
for each choice of $\lambda_1$ and  $\lambda_2$.

Summing over gluon polarizations, one has that:
\begin{eqnarray}
&& \sum_{\lambda_1,\lambda_2} 1 = 4,
%\\ &&
\qquad\qquad\quad
\sum_{\lambda_1,\lambda_2} (\varepsilon_1 \cdot \varepsilon_2)^2 = 2,
\nonumber \\
&&\sum_{\lambda_1,\lambda_2}
(\varepsilon_1 \cdot \varepsilon_2)
(k_1 \cdot \varepsilon_1)
(k_1 \cdot \varepsilon_2) =
m_{\tilde g}^2 - \frac{(t - m_{\tilde g}^2)(u - m_{\tilde g}^2)}{s},
\\
&&\sum_{\lambda_1,\lambda_2}
(k_1 \cdot \varepsilon_1)^2
(k_1 \cdot \varepsilon_2)^2 =
\left (
m_{\tilde g}^2 - \frac{(t - m_{\tilde g}^2)(u - m_{\tilde g}^2)}{s} \right )^2 .
\end{eqnarray}

Following the procedure adopted in the Section~\ref{qqbsgsg} and
taking advantage of the following expressions \cite{dreiner1}:
\begin{eqnarray}
&&
\sum_{\rm colors} G_s^2 = \frac{72 g_s^4}{s^2} ,
\qquad\qquad\qquad\qquad\,\,\,
\sum_{\rm colors} G_t^2 = \frac{72 g_s^4}{(t - m^2_{\tilde g})^2} ,
\\
&&
\sum_{\rm colors} G_u^2 = \frac{72 g_s^4}{(u - m^2_{\tilde g})^2} ,
\qquad\qquad\qquad
\sum_{\rm colors} G_s G_t = \frac{36 g_s^4}{s(t - m^2_{\tilde g})} ,
\\
&&
\sum_{\rm colors} G_s G_u = -\frac{36 g_s^4}{s(u - m^2_{\tilde g})} ,
\qquad\qquad
\sum_{\rm colors} G_t G_u =
\frac{36 g_s^4}{(t - m^2_{\tilde g})(u - m^2_{\tilde g})} , 
\end{eqnarray}
we come to the following differential cross section for the case under study,
which we have also checked with FeynCalc:
\begin{eqnarray}
&&\frac{d \sigma}{d \hat{t}}(gg \to \tilde{g}\tilde{g})= \frac{9 \pi \alpha^{2}_{s}}{4 \hat{s}^{2}} 
\left \{ \frac{2(m^{2}_{\tilde{g}}- \hat{t})(m^{2}_{\tilde{g}}- \hat{u})}{\hat{s}^{2}}+
\frac{m^{2}_{\tilde{g}}( \hat{s}-4m^{2}_{\tilde{g}})}{(m^{2}_{\tilde{g}}- \hat{t})(m^{2}_{\tilde{g}}- \hat{u})}
\right. \nonumber \\ &+& \left. 
\frac{(m^{2}_{\tilde{g}}- \hat{t})(m^{2}_{\tilde{g}}- \hat{u})+2m^{2}_{\tilde{g}}(m^{2}_{\tilde{g}}+ \hat{t})}{(m^{2}_{\tilde{g}}- \hat{t})^{2}} 
+
\frac{(m^{2}_{\tilde{g}}- \hat{t})(m^{2}_{\tilde{g}}- \hat{u})+2m^{2}_{\tilde{g}}(m^{2}_{\tilde{g}}+ \hat{u})}{(m^{2}_{\tilde{g}}- \hat{u})^{2}} 
\right. \nonumber \\ &+& \left.
\frac{(m^{2}_{\tilde{g}}- \hat{t})(m^{2}_{\tilde{g}}- \hat{u})+m^{2}_{\tilde{g}}( \hat{u}- \hat{t})}{\hat{s}(m^{2}_{\tilde{g}}- \hat{t})}+
\frac{(m^{2}_{\tilde{g}}- \hat{t})(m^{2}_{\tilde{g}}- \hat{u})+m^{2}_{\tilde{g}}( \hat{u}- \hat{t})}{\hat{s}(m^{2}_{\tilde{g}}- \hat{u})} \right \}.
\label{formula2}
\end{eqnarray}
Once again, the result agrees the expressions in Refs.~\onlinecite{tata,Dawson}.

\subsection{Subprocess  $qg \to \tilde{q}\tilde{g}$.}
\label{qgsqsg}

\begin{figure}[ht]
\begin{center}
\vglue -0.009cm %\includegraphics{qgsqsgcs.eps}
\mbox{\epsfig{file=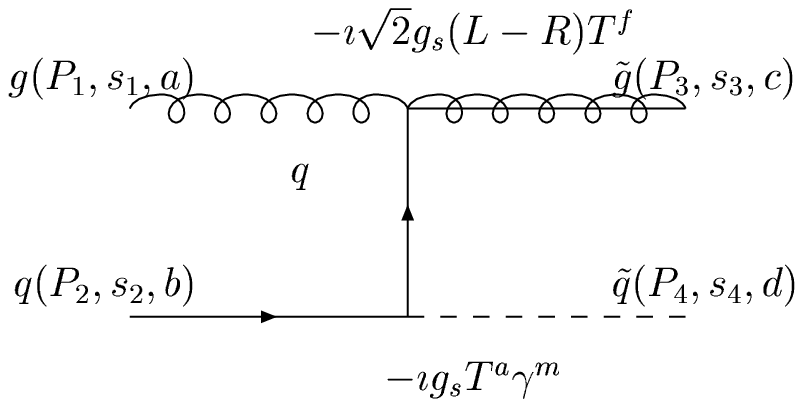,width=0.4\textwidth,angle=0}}
\vglue -0.009cm %\includegraphics{qgsqsgct.eps}
\mbox{\epsfig{file=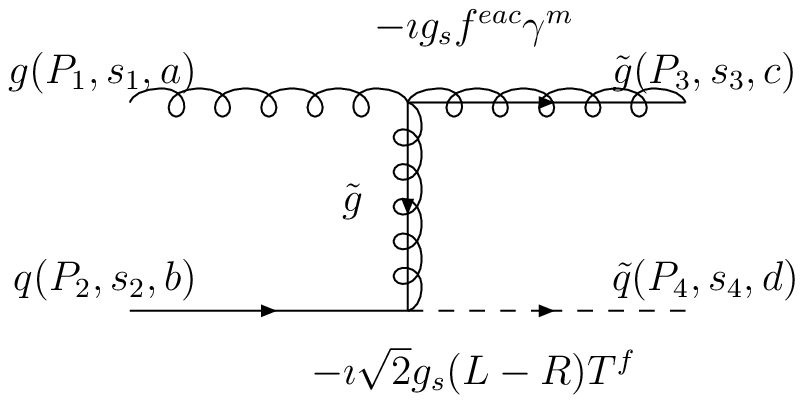,width=0.4\textwidth,angle=0}}
\vglue -0.009cm %\includegraphics{qgsqsgcu.eps}
\mbox{\epsfig{file=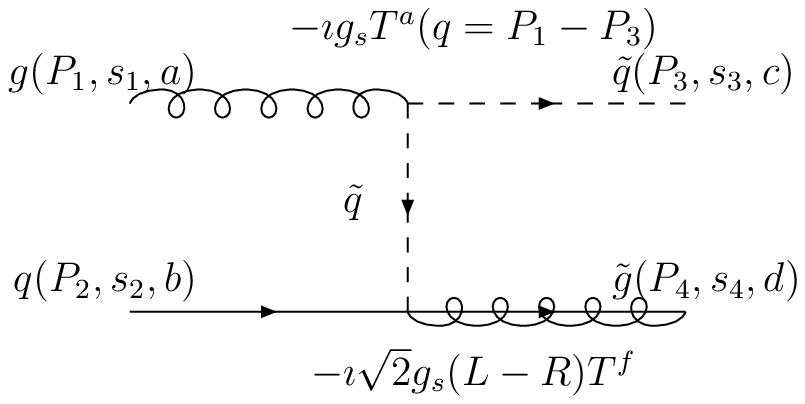,width=0.4\textwidth,angle=0}}  
\end{center}
\caption{Feynman diagrams for squark--gluino production. The arrows
  indicate the fermion flow.}
\label{qgsqsgcs}
\end{figure}

The Feynman diagrams for gluino production from Compton
scattering $qg$ are shown in Fig.~\ref{qgsqsgcs}.

The Feynman amplitudes are given by the equalities
\begin{eqnarray}
{\cal M}_{s}&=&(- \imath \sqrt{2} g_{s}(L-R))
\left( w^{\dagger}(d)T^{f} w(c)u(P_{3},s_{3}) \right)
\left( \frac{\imath ( \not\hspace{-.1ex}{q}+m_{\tilde{g}})}{s}\delta^{fe} \right)
(- \imath g_{s}f^{eac}) \cdot \nonumber \\
&\cdot&
\left( \Omega^{\dagger}(a) \bar{u}(P_{1},s_{1}))T^{e}\gamma^{m}\epsilon^{n}(P_{2})a^{e}(b) \right)\,, \nonumber \\
{\cal M}_{t}&=&(- \imath \sqrt{2} g_{s}(L-R))
\left( w^{\dagger}(d)T^{f} w(b)u(P_{2},s_{2}) \right)
\left( \frac{\imath ( \not\hspace{-.1ex}{q}+m_{\tilde{g}})}{t-m^{2}_{\tilde{g}}}\delta^{fe} \right)
(- \imath g_{s}f^{eac}) \cdot \nonumber \\
&\cdot&
\left( \epsilon^{\mu}(P_{1})a^{e}(a) \gamma^{\mu}u(P_{3},s_{3}))\Omega(c) \right)\,, \nonumber \\
{\cal M}_{u}&=&(- \imath \sqrt{2} g_{s}(L-R))
\left( w^{\dagger}(d)T^{f} w(b)u(P_{3},s_{3}) \right)
\left( \frac{\imath ( \not\hspace{-.1ex}{q}+m_{\tilde{g}})}{u-m^{2}_{\tilde{g}}}\delta^{fe} \right)
(- \imath g_{s}f^{eac}) \cdot \nonumber \\
&\cdot&
\left( \epsilon^{\mu}(P_{1})a^{e}(a) \gamma^{\mu}u(P_{4},s_{4}))\Omega(c) \right).
\label{ampqgsqsgcu}
\end{eqnarray}

The differential cross section for the Compton-like subprocesses is then given by
\begin{eqnarray}
\frac{d \sigma}{d \hat{t}}(qg \to \tilde{q}\tilde{g})&=& \frac{\pi \alpha^{2}_{s}}{24 \hat{s}^{2}} 
\left \{ \left[ \frac{\frac{16}{3}( \hat{s}^{2}+(m^{2}_{\tilde{q}}- \hat{u})^{2})+ \frac{4}{3}\hat{s}(M^{2}_{\tilde{q}}- \hat{u})}
{\hat{s}(m^{2}_{\tilde{g}}- \hat{t})(m^{2}_{\tilde{g}}- \hat{u})} \right] \right. \nonumber \\
&\times& \left. \left( (m^{2}_{\tilde{g}}- \hat{u})^{2}+(M^{2}_{\tilde{q}}-m^{2}_{\tilde{g}})^{2}+ 
\frac{2 \hat{s}m^{2}_{\tilde{g}}(M^{2}_{\tilde{q}}-m^{2}_{\tilde{g}})}{(m^{2}_{\tilde{g}}- \hat{t})} \right) \right \}, 
\label{formula3}
\end{eqnarray}
in agreement with Refs.~\onlinecite{tata,Dawson}.

The total cross section for gluino production can be obtained by
adding Eqs.~(\ref{formula1}), (\ref{formula2}), and (\ref{formula3})
and integrating the resulting equality over phase space, which yields
the results in Refs.~\onlinecite{Dawson,Zerwas}. We have used these
expressions to find the results presented in
Refs.~\onlinecite{Mariotto:2008zt,danusa,danusa2}.

The central purpose of the Large Hadron Collider (LHC) \cite{lhc}, which
is already running and soon will be fully operative with 14 TeV energy,
is to find the Higgs particle. That discovery may either confirm the
Standard Model (SM) or open new windows towards new physics. This
machine will also study collisions involving nuclei-pA
(proton-nucleus, $\sqrt{s}=8.8\, TeV$) and AA (nucleus-nucleus,
$\sqrt{s}=5.5\, TeV$) LHC modes. Results for gluino production in the
pA and AA modes were presented for the first time in
Refs.~\onlinecite{danusa,danusa2}.

Before presenting our numerical results on gluino
production at the LHC, we recall that gluon fluxes and
large color factors make gluon-gluon ($gg$) fusion contributions
dominant at LHC energies if $m_{\tilde{g}},M_{\tilde{q}} \leq 1$ TeV,
while reactions involving valence quarks dominate gluino production at
the Tevatron in the allowed mass range. The rate of gluino pair
production is maximized for $m_{\tilde{g}} \simeq
M_{\tilde{q}}$ \cite{dress}.

\begin{figure}[tb]
\begin{center}
\epsfig{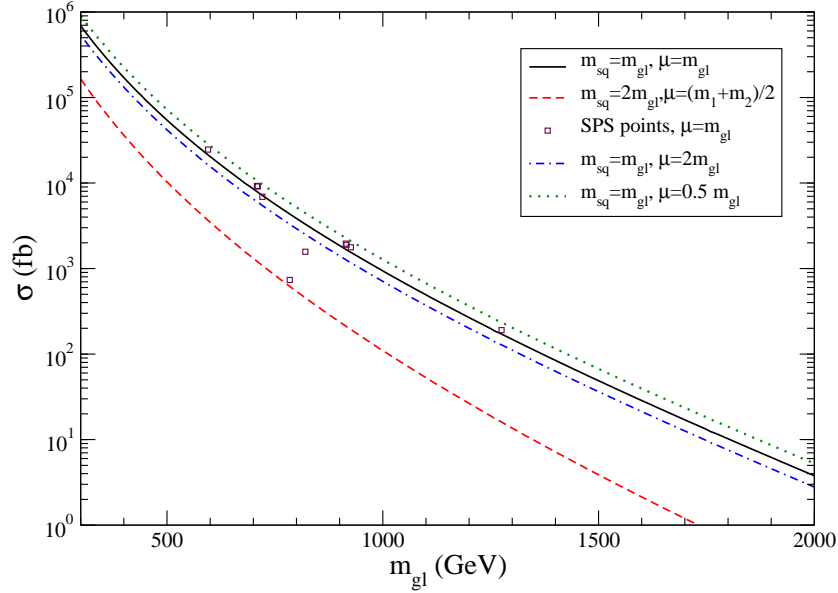}
\end{center}
\caption{Total LO cross section for gluino production at the LHC as a
  function of the gluino masses. Parton densities: CTEQ6L, with two
  assumptions on the squark masses and choices of the hard scale
  (curves). For $m_{\tilde{q}}=m_{\tilde{g}}$ the sensitivity to the
  hard scale is also presented . The open squares show the numerical
  results for the SPS points (see Table~\ref{tab:tmasses}).}
\label{fig:sigtot}
\end{figure}

Figure~\ref{fig:sigtot} shows the LO QCD total cross section for
gluino production at the LHC as a function of the gluino masses.  The
continuous curves display the cross sections calculated for the CTEQ6L
parton densities \cite{Pumplin:2002vw}, with the indicated assumptions
on the squark masses and choices of the hard scale. For
$m_{\tilde{q}}=m_{\tilde{g}}$, the solid, dotted, and dash-dotted
curves monitor the sensitivity to the hard scale.  Our curves are
qualitatively equivalent to the ones in Chapter~12 of
Ref.~\onlinecite{tata}, based on the CTEQ5L parton distribution.

Since the $pp$ CM energy $\sqrt{s}$ =14 TeV is several times larger
than the expected gluino and squark masses, these particles may well
be produced and detected at the LHC. The expected luminosity for the
full LHC performance is ${\cal{L}}\approx
10^{34}$\,cm$^{-2}$\,s$^{-1}$, which is equivalent to $100$~fb$^{-1}$, assuming a full
LHC year of $10^7$\,s. Considering the SPS1a scenarious (lightest
gluino of mass 595.2~GeV), roughly $2\cdot
10^6$ gluinos will be produced, while in the SPS9 scenarious (heavier gluino of mass
1275.2~GeV), $1.8 \cdot 10^4$ gluinos will be produced, according to
our Fig.~\ref{fig:sigtot}. Fore more realistic estimates, the NLO
corrections would increase the cross sections for the various
processes by a factor smaller than two. The above estimates therefore
define a lower limit for the cross section and for the number of produced gluinos.

\section{Conclusion}
\label{sec:conclusion}
We started this article with a very brief review of the early
phenomenology of photinos and gluinos to present our motivation for
calculating the cross section for the production of both particles.
We then highlighted certain difficulties in dealing with Majorana
fermions like we treat Dirac fermions. One difficulty comes from the
ambiguity in the definition of the internal propagators of the
fermions; the other comes from the relative sign between the
amplitudes.  After that we reviewed one method yielding the Feynman
amplitudes when dealing with Majorana fermions, which makes the
calculations as simple as the procedure for Dirac fermions.  This
method is based on a well-defined fermion flow, and yields vertices
equations without explicit charge-conjugation matrices. As
illustrations, we have presented examples showing how to calculate the
photino and gluino production in pp collisions. We expect this review
to help researchers who are new to the field of supersymmetric
extensions of the standard model.

\vbox{
\begin{center}
{\bf Acknowledgments} 
\end{center}
This work was supported by CNPq, DBE supported by Master quota of CNPq
from IF-UFRGS. We are grateful to Pierre Fayet, for several
interesting information about photino and gluino phenomenology. One of
us (MCR) is grateful to Howard E. Haber for pointing out the reference
\cite{dreiner1}, and to J. W. F. Valle for several interesting
comments on leptogenesis.  }


\begin{thebibliography}{99}
\bibitem{gl} Yu. A. Gol'fand and E.P. Likhtman, 
{\sl ZhETF Pis. Red.}{\bf 13}, 452, (1971) [{\sl JETP Lett.}{\bf 13}, 323, (1971)].
\bibitem{va} D.V. Volkov and V.P. Akulov, 
{\sl Phys. Lett.}{\bf B46}, 109, (1973).
\bibitem{wz} J. Wess and B. Zumino, {\sl Nucl. Phys.}{\bf B70}, 39, (1974);
{\sl Phys. Lett.}{\bf B49}, 52, (1974); {\sl Nucl. Phys.}{\bf  B78}, 1, (1974).

\bibitem{Dirac} P.A.M. Dirac, {\sl Proc. Royal Soc. A}{\bf 117}, 610, (1928);
{\bf 118}, 351, (1928).

\bibitem{Majorana} E. Majorana, {\sl Nuovo Cim.}{\bf 14}, 171, (1937).

\bibitem{wb}J. Wess and J. Bagger, {\it Supersymmetry and Supergravity}
Second Edition, Princeton University Press, Princeton NJ, (1992).
\bibitem {dress}M. Drees, R. M. Godbole and P. Royr, {\it Theory and
Phenomenology of Sparticles} First Edition, World Scientific Publishing Co. Pte. Ltd., Singapore, (2004).
\bibitem{MullerKirsten:1986cw}H. J. W. M\"uller-Kirsten and A. Wiedemann, {\it SUPERSYMMETRY: AN INTRODUCTION WITH CONCEPTUAL AND CALCULATIONAL DETAILS}, 
Second Edition, World Scientific Publishing Co. Pte. Ltd., Singapore, (2010).
  %%CITATION = PRINT-86-0955-KAISERSLAUTERN-;%%
  
\bibitem{bailin} D. Bailin and A. Love, {\it Supersymmetric Gauge Field Theory
and String Theory}, First Edition, Institute of Physics Publishing, Bristol UK, (1994).
\bibitem{tata}H. Baer and X. Tata, {\it Weak Scale Supersymmetry},
First Edition, Cambridge University Press, United Kindom, (2006).
\bibitem{ait1}I. Aitchison, {\it Supersymmetry in Particle Physics: An Elementary 
Introduction}, First Edition, Cambridge University Press, United Kindom, (2007).
\bibitem{Srednicki:2004hg}M.~Srednicki, {\it Quantum field theory}, Fourth Edition, Cambridge University 
Press, United Kindom, (2010) and also avaliable at arXiv:hep-th/0409035 and arXiv:hep-th/0409036.




\bibitem{R} P. Fayet, {\sl Nucl. Phys.}{\bf  B90}, 104, (1975).
\bibitem{ssm} P. Fayet, {\sl Phys. Lett.}{\bf B64}, 159, (1976); {\bf B69}, 489, (1977).
\bibitem{grav} P. Fayet, {\sl Phys. Lett.}{\bf B70}, 461, (1977).
\bibitem{Fayet:2001xk}P. Fayet, %``About the origins of the supersymmetric standard model,''
{\sl Nucl. Phys. Proc. Suppl.}{\bf 101}, 81, (2001) (Also in *Minneapolis 2000, 30 years of supersymmetry* 81-98). 
\bibitem{Rodriguez:2009cd} M. C. Rodriguez, %``History of Supersymmetric Extensions of the Standard Model,''
{\sl Int. J. Mod. Phys.}{\bf A25}, 1091, (2010).
\bibitem{Goldberg:1983nd} H. Goldberg,%``Constraint on the photino mass from cosmology,''
{\sl Phys. Rev. Lett.}{\bf 50}, 1419, (1983).
\bibitem{Fayet:1979yb} P. Fayet, %``Scattering Cross-Sections Of The Photino And The Goldstino (Gravitino) On Matter,''
{\sl Phys. Lett.}  {\bf B86}, 272, (1979).

\bibitem{vdWaerden1} B.L. van der Waerden, {\sl Nachrichten Akad. Wiss. G\"ottingen,
Math.-Physik. Kl.}, 100, (1929).


\bibitem{Haber:1994pe}H.E.Haber, %``Spin formalism and applications to new physics searches,''
arXiv:hep-ph/9405376.
\bibitem{Martin:tasi}S. P. Martin, %``TASI 2011 lectures notes: two-component fermion notation and supersymmetry,''
arXiv:1205.4076.
\bibitem{dreiner1}H. K. Dreiner, H. E. Haber and S. P. Martin,
%``Two-component spinor techniques and Feynman rules for quantum field theory and supersymmetry,''
{\sl Phys. Rept.}{\bf 494}, 1, (2010).

\bibitem{bhabha}H. J. Bhabha, {\sl Proc. Roy. Soc.}{\bf A154}, 195, (1935).

\bibitem{ff} G.R. Farrar and P. Fayet, {\sl Phys. Lett.} {\bf B76}, 575, (1978).
\bibitem{ff2} G.R. Farrar and P. Fayet, {\sl Phys. Lett.} {\bf B79}, 442, (1978).
\bibitem{Cremmer:1982vy}E. Cremmer, P. Fayet and L. Girardello,%``Gravity Induced Supersymmetry Breaking And Low-Energy Mass Spectrum,'' 
{\sl Phys.\ Lett.}{\bf B122}, 41, (1983).



%\bibitem{Chung:2003fi} D. J. H. Chung, L. L. Everett, G. L. Kane, S. F. King, J. D. Lykken and L. T. Wang, {\sl Phys.Rept.}{\bf 407}, 1, (2005).


\bibitem{Fayet:1982ky} P. Fayet, %``Radiative Production Of Gravitinos And Photinos In E+ E- Annihilation,''
{\sl Phys. Lett.}{\bf B117}, 460, (1982).
\bibitem{kane} H.E. Haber and G.L. Kane, {\sl Phys. Rep.}{\bf 117}, 75, (1985).
\bibitem{Dawson}S. Dawson, E. Eichten and C. Quigg, %``Search For Supersymmetric Particles In Hadron - Hadron Collisions,''
{\sl Phys. Rev.}{\bf D31}, 1581, (1985).

\bibitem{Kobayashi:1984wu}T. Kobayashi and M. Kuroda, %``Photino Mass And Gamma Photino Photino Production In E+ E- Annihilation,''
{\sl Phys. Lett.}{\bf B139}, 208, (1984).
\bibitem{Grassie:1983kq}K. Grassie and P. N. Pandita, %``Production Of Photinos In E+ E- $\to$ Gamma Photino Photino,''
{\sl Phys. Rev.}{\bf D30}, 22, (1984).
\bibitem{Ware:1984kq}J. D. Ware and M. E. Machacek,  %``Photino-Photino - Photon Production In E+ E- Annihilation,''
{\sl Phys. Lett.}{\bf B142}, 300, (1984).
\bibitem{Bento:1985in} L. Bento, J. C. Romao and A. Barroso, %``E+ E- $\to$ Gamma + Missing Neutrals: Neutrino Versus Photino Production,''
{\sl Phys. Rev.}{\bf D33}, 1488, (1986).
\bibitem{pandita}R. Basu, P. N. Pandita and C. Sharma, %``Radiative neutralino production in low energy supersymmetric models"
{\sl Phys. Rev.}{\bf D77}, 115009, (2008).

\bibitem{Berezinsky:1986ty}V. S. Berezinsky, E. V. Bugaev and E. S. Zaslavskaya, %``On High-Energy Cosmic Photinos,''
{\sl Nucl. Phys.}{\bf B272}, 193, (1986).


\bibitem{Baer:1986au}H. Baer, V. D. Barger, D. Karatas and X. Tata, %``Detecting Gluinos at Hadron Supercolliders,''
{\sl Phys. Rev.}{\bf D36}, 96, (1987).
\bibitem{Baer:1990sc}H. Baer, X. Tata and J. Woodside, %``PHENOMENOLOGY OF GLUINO DECAYS VIA LOOPS AND TOP QUARK YUKAWA COUPLING,''
{\sl Phys. Rev.}{\bf D42}, 1568, (1990).
\bibitem{Haber:1983fc}H. E. Haber and G. L. Kane, %``Gluino Decays And Experimental Signatures,''
{\sl Nucl. Phys.}{\bf B232}, 333, (1984).
\bibitem{Ma:1988ns}E. Ma and G. G. Wong, %``TWO-BODY RADIATIVE GLUINO DECAYS,''
{\sl Mod. Phys. Lett.} {\bf A3}, 1561, (1988).
\bibitem{Barbieri:1987ed}R. Barbieri, G. Gamberini, G. F. Giudice and G. Ridolfi, %``CONSTRAINING SUPERGRAVITY MODELS FROM GLUINO PRODUCTION", 
{\sl Nucl. Phys.}{\bf B301}, 15, (1988).

\bibitem{resultssusysearches} G.~Aad {\it et al.}  [ATLAS Collaboration],
  %``Search for squarks and gluinos using final states with jets and missing transverse momentum with the ATLAS detector in sqrt(s) = 7 TeV proton-proton collisions,''
    Phys.\ Lett.\ {\bf B701}, 186 (2011); Phys.\ Lett.\ {\bf B710}, 67 (2012); arXiv:1208.3144 [hep-ex].
\bibitem{sps1}B.C. Allanach {\it et al}, {\sl Eur.Phys.J.}{\bf C25}, 113, (2002).
\bibitem{sps2}Nabil Ghodbane and Hans-Ulrich Martyn, hep-ph/0201233.
\bibitem{sps} http://spa.desy.de/spa/

\bibitem{sg}S.J.~L.~Rosner,
  %``Resource letter: The Standard model and beyond,''
{\sl Am. J. Phys.} {\bf 71}, 302, (2003); 
A.~S.~Kronfeld and C.~Quigg,
  %``Resource Letter: Quantum Chromodynamics,''
{\sl Am. J. Phys.}{\bf 78}, 1081, (2010).

\bibitem{denner}A. Denner, H. Eck, O. Hahn and J. K\"ublbeck, {\sl Nucl. Phys.}{\bf B387}, 467, (1992).
\bibitem{denner1}A. Denner, H. Eck, O. Hahn and J. K\"ublbeck, {\sl Phys. Lett.}{\bf B387}, 278, (1992).

\bibitem{gates}E.I. Gates, and K. L. Kowalski, {\sl Phys. Rev.}{\bf D45}, 1693, (1992).
\bibitem{Hahn:2000kx}T. Hahn, %``Generating Feynman diagrams and amplitudes with FeynArts 3,''
{\sl Comput. Phys. Commun.}{\bf 140}, 418, (2001).
\bibitem{Bartl:1989ms}A. Bartl, H. Fraas, W. Majerotto and N. Oshimo, %``The Neutralino Mass Matrix In The Minimal Supersymmetric Model,''
{\sl Phys. Rev.}{\bf D40}, 1594, (1989).
\bibitem{Carena:1986jp}M. S. Carena and C. E. M. Wagner, %``NEUTRALINO PRODUCTION IN e+ e- ANNIHILATION,''
{\sl Phys. Lett.} {\bf B195}, 599, (1987).


\bibitem{Olive:1994qq}K. A. Olive and S. Rudaz, %``Light stops in the MSSM: Implications for photino dark matter and top quark decay,''
{\sl Phys. Lett.}{\bf B340}, 74, (1994).
\bibitem{ilc}http://www.linearcollider.org/

\bibitem{Brau:2012zz}
  J.~E.~Brau, (ed.) {\it et al.},
  ``International Linear Collider Physics and detectors: 2011 Status
Report,''
  CERN-LCD-NOTE-2011-038.

\bibitem{Zerwas} W. Beenakker, R. H\"opker, M. Spira and P.M. Zerwas, {\sl Nucl. Phys.}{\bf B492}, 51, (1997).

\bibitem{Mariotto:2008zt}C. B. Mariotto and M. C. Rodriguez, {\sl Braz. J. Phys.}{\bf 38}, 503, (2008) and 
%``Gluino production in some supersymmetric models at the LHC", 
arXiv:0805.2094 [hep-ph].
\bibitem{danusa}D.B. Espindola, "Produ\c c\~ao de Fotinos e Glu\'\i nos nas Extens\~oes Supersim\'etricas 
da Eletrodin\^amica Qu\^antica e da Cromodin\^amica Qu\^antica", Master thesys defended at 03/18/2010.

\bibitem{plehn}W.~Beenakker, M. Kr\"amer, T. Plehn, M. Spira and P. M. Zerwas, %``Stop production at hadron colliders,''
{\sl Nucl. Phys.}{\bf B515}, 3, (1998).

\bibitem{Hahn:2001rv} T. Hahn and C. Schappacher, %``The implementation of the minimal supersymmetric standard model in FeynArts and FormCalc,''
{\sl Comput. Phys. Commun.}  {\bf 143}, 54, (2002).


\bibitem{danusa2}C. Brenner Mariotto, D. B. Espindola and M. C. Rodriguez, 
%``Gluino production in ultrarelativistic heavy ion collisions and nuclear shadowing'', 
{\sl Phys.\ Rev.}\  {\bf C83}, 064902 (2011).
\bibitem{lhc}http://lhc.web.cern.ch/lhc/
\bibitem{Pumplin:2002vw}J. Pumplin, D. R. Stump, J. Huston, H. L. Lai, P. Nadolsky and W. K. Tung,
{\sl JHEP}{\bf 0207}, 012, (2002).



\end{thebibliography}
\end{document}